\documentclass[11pt,letterpaper]{article}
\usepackage{graphicx}% Include figure files
\usepackage{bm}% bold math
\usepackage{amssymb}
\usepackage{color}
\usepackage{amsmath}
\usepackage[numbers]{natbib}
\usepackage{enumerate}
\usepackage{amstext}
\usepackage{footmisc}
\usepackage{latexsym}
\usepackage[usenames,dvipsnames]{xcolor}
\usepackage[colorlinks=true,citecolor=Cerulean,linkcolor=RubineRed,urlcolor=Cerulean]{hyperref}
\usepackage{multibib}
\usepackage{ fixfoot}
\usepackage{fancyhdr}
\newcommand{\pac}[1]{ \left\{ #1 \right\} }
\newcommand{\pap}[1]{\left( #1 \right)}
\newcommand{\pas}[1]{\left[#1 \right]}
\newcommand{\paV}[1]{\left\Vert #1 \right\Vert}

\newcommand{\ket}[1]{ \left| #1 \rangle\right.}
\newcommand{\bra}[1]{  \left.\langle #1  \right|}

\newcommand{\angstrom}{\textup{\AA}}
\DeclareFixedFootnote{\rep}{Electronic address: \href{mailto:fernandojavier.gomez@iff.csic.es}{fernandojavier.gomez@iff.csic.es}}

\def\tr{\rm{Tr}}

\newcommand{\beq}{\begin{equation}}
\newcommand{\eeq}{\end{equation}}
\newcommand{\beqa}{\begin{eqnarray}}
\newcommand{\eeqa}{\end{eqnarray}}

\begin{document}
\title{{\bf Energy transfer in $N$-component nanosystems enhanced by pulse-driven vibronic many-body entanglement}}
\author{ Fernando J. G\'omez-Ruiz\href{https://orcid.org/0000-0002-1855-0671}{\includegraphics[scale=0.45]{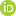}}\,$^{1,2}$, Oscar~L. Acevedo$^{3}$, Ferney J. Rodr\'iguez\href{https://orcid.org/0000-0001-5383-4218}{\includegraphics[scale=0.45]{orcid}}\,$^{4}$, \\
Luis Quiroga\href{https://orcid.org/0000-0003-2235-3344}{\includegraphics[scale=0.45]{orcid}}\,$^{4}$, and Neil~F. Johnson$^{5,\dagger}$}
\date{}
\maketitle
\vspace{-1cm}
\begin{center}
$^{1}${\it Departamento de F\'isica Te\'orica, At\'omica y \'Optica, Universidad de Valladolid, 47011 Valladolid, Spain}\\
$^{2}${\it Instituto de F\'isica Fundamental IFF-CSIC, Calle Serrano 113b, Madrid 28006, Spain}\\
$^{3}${\it Escuela de Ciencias B{\'a}sicas, Instituci{\'o}n Universitaria Polit{\'e}cnico Grancolombiano, Bogot{\'a} D.C. 110231, Colombia}\\
$^{4}${\it Departamento de F{\'i}sica, Universidad de los Andes, A.A. 4976, Bogot\'a D. C., Colombia}\\
$^{5}${\it Physics Department,  George Washington University, Washington D.C. 20052, U.S.A.}\\
$^{\dagger}${\it Corresponding Author:} \href{mailto:neiljohnson@gwu.edu}{neiljohnson@gwu.edu}
\end{center}
\begin{abstract}
The processing of energy by transfer and redistribution, plays a key role in the evolution of dynamical systems. At the ultrasmall and ultrafast scale of nanosystems, quantum coherence could in principle also play a role and has been reported in many pulse-driven nanosystems (e.g. quantum dots and even the microscopic Light-Harvesting Complex II  (LHC-II) aggregate). Typical theoretical analyses cannot easily be scaled to describe these general $N$-component nanosystems; they do not treat the pulse dynamically; and they approximate memory effects. Here our aim is to shed light on what new physics might arise beyond these approximations. We adopt a purposely minimal model such that the time-dependence of the pulse is included explicitly in the Hamiltonian. This simple model generates complex dynamics: specifically, pulses of intermediate duration generate highly entangled vibronic (i.e. electronic-vibrational) states that spread multiple excitons -- and hence energy -- maximally within the system. Subsequent pulses can then act on such entangled states to efficiently channel subsequent energy capture. The underlying pulse-generated vibronic entanglement increases in strength and robustness as $N$ increases.
\end{abstract}
\section*{INTRODUCTION}
There has been an exciting development over the past decade concerning chemical, biophysical and physical systems driven by short bursts of energy (e.g. laser) in which the subsequent energy processing involves coherence phenomena~\cite{Nishida_NatCom_22, Jianshu_SciEDv2020, Lambert2013, Scholes_1, OReilly2014_5, Plenio_23}. Certain aspects of this have been reported as arising from quantum-mechanical interference interactions between the electronic and vibrational constituents within the system. Examples of such systems include photosynthetic complexes~\cite{Donatas2022,Olaya_PRB08,Thila,Fujihashi_3,Chin_4,Scholes_11,Fleming_13,Collini_15,Rozzi_17,Tiwari_22,Schlau_26,Scholes_27,Scholes_30}, metal surfaces~\cite{Hainer2021,Reutzel_PRX19,Bittner_39}, molecular magnets~\cite{Liedy2020,Rogers_23,Canton_23}, biochemical control~\cite{Paulus2020,Coccia_2019,Gaynor2019,Wang_16and1,Liu_38}, organic devices~\cite{Bian2020,Dubin_10,Collini_14}, and more general nanostructures~\cite{Cassette_2,Scholes_25}. Understanding the impact of a strong, time-dependent perturbation (e.g. pulse) on  many-body quantum mechanical correlations, is a necessary step in fully understanding how energy is processed in time in such systems -- and its consequences for energy transfer. However, typical  theoretical analyses of such systems cannot easily be scaled to the large numbers $N$ of components in either a naturally occurring real sample or an artificially made device, and tend to average over or truncate memory effects. \\
\\
The present paper is motivated by this outstanding theoretical challenge of treating the dynamics of the pulse on the same footing as the dynamics of the general $N$ many-body quantum nanostructure system.  The literature on such many-body quantum systems contains various theoretical methods which have been extended to include complications such as non-Markovianity, other forms of perturbations, and quantum phenomena like coherence and entanglement. Particularly notable examples include the reduced hierarchical equations of motion theory~\cite{Tanimura_HEOM,Lambert2019}, path-integral Monte Carlo~\cite{Egger_PRB94, Cao_96}, the quasi-adiabatic path-integral algorithm~\cite{Makri_95,Makri2_95}, and the time-dependent Davydov ansatz~\cite{Zhao_23}. These methods typically entail a significant computational overhead and their complicated details may obscure the interplay between various competing physics effects. For a comprehensive review of these numerical methods, we refer to Ref.~\cite{Tanimura_HEOM,Zhao_23}.\\
\\
There is an important caveat to our aims and hence this paper: we are not aiming at a detailed understanding of a specific real-world system with highly complex chemistry such as the light-harvesting complex LHC-II, but instead our focus is on advancing the basic physics understanding of time-dependent response in a many-body quantum system driven by a pulse of arbitrary strength. Since the problem is impossible to solve exactly for arbitrarily strong pulses in a system with complex chemistry such as LHC-II, we adopt the well-known approach of physicists of focusing entirely on a minimal  model. While such a model is effectively a `toy' for theoreticians like us to play with, this approach is well-established in Physics where seemingly simple models like the Ising model are employed to mimic collective behavior in systems with exceedingly complex chemical properties. Specifically, we include the time-dependent pulse as part of a so-called Dicke-like Hamiltonian of $N$ two-level systems -- not as a perturbation -- and then we solve this model's time-dependence exactly. Though this sacrifices most of the chemical details which could ultimately prove to be important for a specific real-world system (e.g. LHC-II), we show that doing so has the advantage of making the model exactly solvable numerically for any $N$ without making any assumptions about the pulse being less than a certain strength, and without making any approximations concerning memory effects in the system's quantum mechanical evolution. Hence our purposely minimal model simplifies all the complexities of any real-world nanostructure (e.g. chemistry) in order to explore the physical interplay of the dynamics of the pulse and the dynamics of the many-body quantum system with any $N>1$. We make no claims that this model is an accurate description of a complex system such as LHC-II -- however our minimal model's exact numerical solution does reveal new physical collective behavior that should transcend such chemical details and hence could arise in such systems, i.e. we show for the first time that pulses of intermediate duration generate highly entangled vibronic (i.e. electronic-vibrational) states that spread multiple excitons -- and hence energy -- maximally within the system. Subsequent pulses can then act on such entangled states to efficiently channel subsequent energy capture. The underlying pulse-generated vibronic entanglement increases in strength and robustness as $N$ increases. The possibility of such effects in a complex system such as LHC-II at very short timescales, cannot therefore be ruled out.\\
\begin{figure*}[t!]
\begin{center}
\includegraphics[scale=1.2]{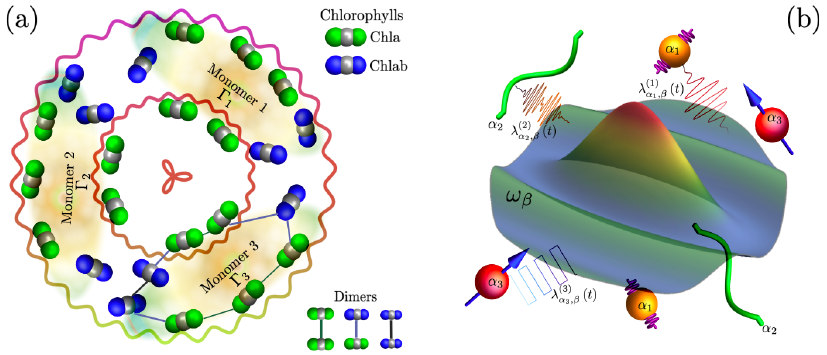}
\caption{\label{fig_1}{\bf Potential applications.} {\bf (a)}  Schematic of the light-harvesting LHC-II system containing various large-scale so-called `monomers', each of which is an aggregate containing chlorophyll dimers \cite{Trimer,OReilly2014_5}. We refer to Refs. \cite{Trimer,OReilly2014_5} for full chemical details. Zooming in, each chlorophyll molecule comprises a central magnesium atom (gray) and two nitrogen atoms~\cite{Trimer}. The nitrogen atoms for Chla (Chlb) are colored green (blue). These chlorophylls are densely packed within a confined space of approximately $\sim 50 \angstrom$. Associated with each of these three `monomers', there are pairs of chlorophylls that can potentially form dimers and that form approximate two-level systems with very similar energy splittings\cite{OReilly2014_5}: this suggests that LHC-II contains $N=3$ chlorophyll dimer two-level systems with similar energy splittings akin to our minimal model\cite{OReilly2014_5}. Specifically, when stripped of all its chemical details, the system is a candidate $N=3$ system, i.e. three chlorophyll dimers with similar energy splittings can hence act approximately as $N=3$ two-level systems. {\bf (b)} Schematic of a more general nanostructure system represented by our minimal model. The overall complex system comprises various types of nanostructure components denoted by $\alpha_i$, immersed within a bosonic bath. Each mode within this bath possesses an elemental frequency $\omega_\beta$. Each nanostructure component $\alpha_i$ can be effectively approximated as a two-level system. Furthermore, each nanostructure component is subject to the influence of a time-dependent pulse represented by $\lambda_{{\alpha_i},\beta}^{(i)}$ (see Eq.~\ref{hdic}).}
\end{center}
\end{figure*}

With this important caveat in mind, we will proceed in this paper to explore  this minimal theoretical physics model and we will show that it has interesting new physics outcomes when solved exactly numerically. We simply {\em speculate} that it may serve as a very simplified model of complex systems such as LHC-II and hence its novel properties and results reveal previously overlooked behaviors in such systems. We will therefore refer to LHC-II in what follows and present it in our figures, in order to explain its fascinating structure to a broader audience and also to act as an illustrative example in order to motivate the general problem. We also hope that doing this will motivate an eventual full-scale calculation, using future computing power, in which all the chemistry details of prior models can somehow be included on the same footing as the time-dependent pulse, so that the full system can be solved exactly as here. Such light-harvesting complexes (LHC-II) are found in green plants (width $\sim 50 \angstrom$) and play a pivotal role in capturing solar energy. Figure~\ref{fig_1} (a) shows a schematic of the LHC-II structure, adapted from Ref. \cite{Trimer}: it consists of an overall trimeric arrangement housing 24 chlorophylls and it is composed by two nitrogen and one magnesium atoms. These chlorophylls are organized into two irregular circular rings, as illustrated in Figure~\ref{fig_1} (a). The inner ring, situated at the core of the trimer, comprises six Chla molecules believed to play a crucial role in energy transfer. The remaining chlorophylls are strategically arranged to facilitate efficient absorption of incident light energy from all directions (for more details see Ref.~\cite{Trimer,Drop,Lambrev}).  Figure~\ref{fig_1} (b) gives a schematic representation of general nanostructure systems represented by our minimal model (Eq. 1) for which LHC-II may be an example. Adopting the simplest scenario, we approximate each individual nanostructure component as a two-level system. The two states $\ket{X_{i}}$ and $\ket{Y_{i}}$, represent the higher and lower energy states for the $i$-th nanostructure component (i.e. for the $i$-th dimer in the case of LHC-II where each dimer is a chlorophyll pair\cite{OReilly2014_5}). In LHC-II which features $N=3$ chlorophyll-pair dimers with similar energy splittings, the wavefunction for each dimer's higher excitonic state $|X_\Gamma\rangle$ is more localized on the site with higher energy, while the lower state $|Y_\Gamma\rangle$ is more localized on the site with lower energy, hence a transition $|X_\Gamma\rangle\rightarrow |Y_\Gamma\rangle$ represents excitonic energy transfer in space from one to the other. This system motivates the focus in this paper on $N=3$, though our main findings hold for more general $N$ (Figure~\ref{fig_1} (b)).\\
\\
Summarizing the rest of the paper, we will proceed to set up and solve numerically the real-time evolution of our minimal model. We will then show that pulses of intermediate duration generate strong $N$-body vibronic entanglements (i.e. between the electronic and vibrational subsystems) which represent a novel quantum-mechanical form of coherence, and which enhance the transfer and subsequent channeling of energy across the system. Our calculations predict that the strength and robustness of these $N$-body vibronic entanglements, increase with $N$. Despite our minimal model's simplicity, it therefore yields novel results that are the first to our knowledge to account for the full many-body entanglement evolution for $N\geq 3$ specifically, the full pulse dynamics and full memory effects. In this way, it avoids the typical previous approximations of small $N$, linear response and truncated memory effects. Again we stress that our intention is not to address the longer timescale mechanisms driving natural photosynthesis in hot, wet environments~\cite{PNASnew}. Instead, our results help deepen understanding of the temporal quantum evolution in pulsed systems of general size $N$ and we speculate that this can shed new light on the early-time kinetics in such real-world open systems, since in this temporal regime the timescale is too short to couple in many of the complex degrees of freedom that will naturally exist in such real-world systems~\cite{Kenrow}.

\section*{OUR MINIMAL MODEL}
As is well known from classical and quantum optics, any incident electromagnetic (light) field ${\vec E}(t)$ generates an internal polarization field ${\vec P}({\vec r},t)$ within the material, given exactly by Maxwell's Equations~\cite{Li,Exc}. Though nonlinear and anisotropic in general, the presence of $\partial^2/{\partial t^2}$ terms for both  means that a pulse in ${\vec E}$ will generate a similar pulse in ${\vec P}$, and hence a pulse in the internal electric field dynamics coupling the electronic and vibrational systems~\cite{Li,Exc}. Since we are not focusing on single-photon phenomena, a similar conclusion follows from a quantum-mechanical starting point~\cite{Johnson,PRBluis}. This helps motivate our minimal model which is a Dicke-like Hamiltonian featuring time-dependent electronic-vibrational coupling $\lambda(t)$ (throughout the manuscript, we employ $\hslash =1$):
\begin{equation}\label{hdic}
{H_N}(t)=\sum_\beta\omega_\beta a_\beta^{\dagger}{a_\beta} + \sum_{i=1}^{N}\sum_{\alpha_i\in i}\frac{\epsilon_{\alpha_i}}{2}{\sigma}_{z,\alpha_i}^{i}  +\sum_\beta  \sum_{i=1}^{N}\sum_{\alpha_i\in i}
\frac{\lambda_{\alpha_i,\beta}^{(i)}(t)}{\sqrt{N}}\left(a_\beta^{\dagger}+{a_\beta}\right){\sigma}_{x,\alpha_i}^{i}
\end{equation}
where $N$ is the number of components that respond to the pulse. We then integrate numerically this Hamiltonian system's time-dependent equation in order to obtain its time-dependent quantum mechanical solutions. Other degrees of freedom that remain inert on such short timescales, can be neglected in this initial study but could be included later on through perturbation theory.  ${\sigma}_{p,\alpha_i}^{i}$ denotes the two-level Pauli operators for excitation $\alpha_i$ on each component $i$ with $p=x,z$. Here, we consider the elemental excitation energy for each nanostructure component $\alpha_i$ denoted by $\epsilon_{\alpha_i}$. The vibrational modes $\beta$ may or may not be localized, and can include relative modes~\cite{Vivek_22}. In this context, the operator $a_{\beta}^{\dagger}$ ($a_{\beta}$) is responsible for generating (eliminating) a photon within the mode $\beta$ with energy $\omega_\beta$. $\lambda_{\alpha_i,\beta}^{i}$ corresponds to an individual time-dependent pulse applied to the nanostructure component $\alpha_i$ within mode $\beta$. Generally, each pulse can assume a unique time-dependent form tailored for experimental control purposes. Though Eq.~\eqref{hdic} obviously sacrifices the chemical details of any given real-world system, it focuses attention on the key physical factors controlling the system's complex quantum evolution. In terms of the speculated application to LHC-II, experimental observations of LHC-II suggest that the dimers exhibit a close resonance between their two excitonic levels and one of the background vibrational modes. Specifically, within the Chlb$_{601}$-Chla$_{602}$ dimers, the hybridized exciton energy-level splitting stands at $\epsilon=667.7$ cm$^{-1}$, while a vibrational mode with $\omega_{\rm vib}=742.0$ cm$^{-1}$ is present~\cite{OReilly2014_5}. Under the resonant conditions considered here, an $\epsilon=\omega=1$ pair is retained. As we show in this paper, this means in principle that many-body states can now be generated that entangle the electronic and vibrational subsystems as a result of the speed of changes in $\lambda(t)$ but {\em without} having to access the strong-coupling regime~\cite{Johnson,Acevedo2014PRL}. Many variants of Eq.~\eqref{hdic} will show similar results due to an established universal dynamical scaling \cite{Acevedo2014PRL} and the fact that they typically generate similar types of phase diagrams and hence have similar collective states in the static $\lambda$ limit~\cite{Lee2007_a,Jarrett_b,Lee_c,Jarrett_d}. Full background to our numerical calculations, including the multiple checks that we performed for convergence, are given in Refs.~~\cite{Acevedo2015NJP,AcevedoPRA2015,GomezEnt2016,GomezJPB2018,Fabio_2020,Fabio_2023}.
%================Results and Discussions  ===============

\section*{RESULTS AND DISCUSSION }
\subsection*{Transfer of energy}
In order to simulate a driving pulse, we consider a time-dependent coupling term in Eq.~\ref{hdic}, $\lambda(t)$, with an up-down linear ramping in which the coupling strength goes from $0\rightarrow 1$ and then symmetrically $1 \rightarrow 0$. The pulse is characterized by a specific ramping velocity $\upsilon$,  where $\upsilon^{-1}$ sets the pulse duration. Hence, the driving pulse is expressed as follows:
\begin{equation}\label{pulse}
\lambda\pap{t}=\begin{cases}
\upsilon t,& 0\leq  t \leq \upsilon^{-1}, \\
2-\upsilon t,& \upsilon^{-1} < t \leq 2\upsilon^{-1}.
\end{cases}
\end{equation}
The particular choice of $\lambda$ given by Eq.~\eqref{pulse}, implies a round-trip protocol. Recently, the round-trip protocols has garnered interest in the context of quantum scaling properties in critical systems~\cite{Tarantelli_PRB22,Franco_PRB23}. We numerically solve the time-dependent Schr\"odinger equation to obtain the instantaneous system state $\ket{\psi\pap{t}}$. \\
\\
\begin{figure*}[t!]
\begin{center}
\includegraphics[scale=0.75]{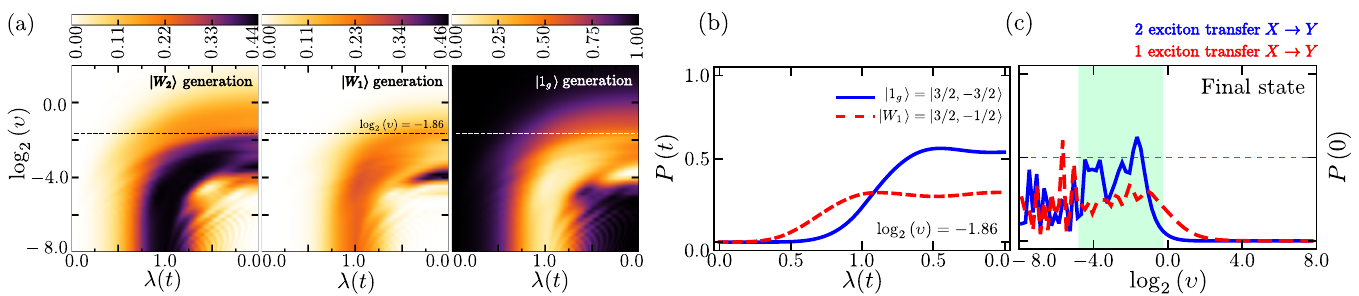}
\caption{\label{fig_2} {\bf Example of $N=3$. (a)} Colors show time-dependent probability profile of generating particular $N=3$ body multi-exciton states (left, middle and right panels), as a function of time during a single up-down pulse (horizontal axis) while the vertical axis shows the logarithm of the ramping velocity $\upsilon$. In all cases the matter initial state is $\ket{1_g}$. The panels {\bf (b)} and {\bf (c)} show the contrast of impact when the starting state is $\ket{1_g}$ or $\ket{W}_2$. {\bf (b)} Time-dependent probability of multi-exciton transfer from $\ket{X_\Gamma}$ to $\ket{Y_\Gamma}$ (i.e. to lower energy chromophore in LHC-II) during pulse of intermediate duration. {\bf (c)} Final state probabilities following pulse, as function of inverse pulse duration $\upsilon$. At intermediate durations (shown as green shaded in {\bf (c)} and explicitly in {\bf (b)}) the transfer and hence energy transport of $2$ excitons has higher final probability than that of $1$ exciton.}
\end{center}
\end{figure*}

We now examine in more depth the illustrative scenario where a single type of component is present $\alpha_i = \alpha$ and there are three of them ($N=3$), inspired by LHC-II with its three chlorophyll dimers of similar energies. We will focus on initially separated matter-vibrational states with zero vibrational excitations.  Within each two-level system, there are two hybridized excitonic states $|X_\Gamma\rangle$ (higher excitonic energy state) and $|Y_\Gamma\rangle$ (lower excitonic energy state). For a simple asymmetric double-well system, akin to the chlorophyll-pair dimers in LHC-II, the wavefunction for the higher dimer state $|X_\Gamma\rangle$ is more localized on the site with higher energy, and the lower dimer state $|Y_\Gamma\rangle$ is more localized on the site with lower energy. Hence a transition $|X_\Gamma\rangle\rightarrow |Y_\Gamma\rangle$ represents a spatial transfer of exciton energy in space from one to another~\cite{OReilly2014_5}. Importantly for light-harvesting, the $|Y_\Gamma\rangle$ wavefunction is localized on the chlorophyll {\em nearest} to an energy exit site. Hence any transition of exciton(s) from state(s) $\{|X_\Gamma\rangle\}$ to $\{|Y_\Gamma\rangle\}$ represents a spatial transfer of their energy toward a site where that energy can be easily exported. The initial states of the entire system including the vibrational modes are as follows: $\ket{\psi_1\pap{0}}=\ket{1_g}\otimes\ket{0}$ and $\ket{\psi_2\pap{0}}=\ket{W_2}\otimes\ket{0}$, respectively. The multi-excitonic states $\ket{1_g}$, $\ket{W_1}$, and $\ket{W_2}$ can be expressed in the conventional Dicke-like manifold basis $\ket{J,J_z}$ as follows: $\ket{1_g}=\ket{3/2,-3/2}$, $\ket{W_1} = \ket{3/2,-1/2}$, and $\ket{W_2} = \ket{3/2,1/2}$, where $J=3/2$ and $J_z$ takes on the values $-3/2$, $-1/2$, and $1/2$. These states can also be expressed in the excitonic basis of the $N=3$ two-level systems as follows, indicating a total of 0, 1 and 2 excitons respectively:
\begin{subequations}
\begin{align}
\ket{1_g}&\equiv\ket{Y_1,Y_2,Y_3},\label{gs_state}\\
\ket{W_1}  &\equiv\frac{1}{\sqrt{3}}\left(\ket{Y_1,Y_2,X_3}+\ket{Y_1,X_2,Y_3}+\ket{X_1,Y_2,Y_3}\right),\label{W1_state}\\
\ket{W_2}&\equiv\frac{1}{\sqrt{3}}\left(\ket{X_1,X_2,Y_3} + \ket{X_1,Y_2,X_3} + \ket{Y_1,X_2,X_3}\right).\label{W2_state}
\end{align}
\end{subequations}
For example in the state $\ket{W_2}$, two matter excitations are shared among the three nanostructure components. We then evaluate the time-dependent probability of finding the system's instantaneous state projected onto a given reference quantum state, pure $\ket{\phi_f}$ or mixed $\rho_f$ states. In general, the fidelity or probability for the instantaneous system's density matrix $\rho\pap{t}=\ket{\psi\pap{t}}\bra{\psi\pap{t}}$ is given by
\begin{equation}\label{fid}
P\pap{t}:=\pas{{\rm Tr}\sqrt{\sqrt{\rho_f}\rho\pap{t}\sqrt{\rho_f}}}^{2}.
\end{equation}
Information about the electronic (matter) subsystem is extracted from the corresponding reduced matrix, $\rho_{e}\pap{t}={\rm Tr}_{v}\pas{\rho\pap{t}}$ where ${\rm Tr}_{v}\pas{\ldots}$ means that the vibrational states have been traced out. In Fig.~\ref{fig_2}(a), we depict the time-dependent probability profile (fidelity) given by direct evaluation of Eq.~\ref{fid} starting with the total state given by $\ket{\psi_1\pap{0}}=\ket{1_g}\otimes\ket{0}$, where the state $\ket{0}$ correspond to zero excitations in the bosonic (i.e. vibrational) mode. In each panel, the corresponding reference state is shown in the upper part of the frame. In Fig.~\ref{fig_2}(b), we have considered a scenario in which the complex system is initially prepared in the state $\ket{\psi_2 (0)}=\ket{W_2}\otimes \ket{0}$. We have computed the probabilities of the matter subsystem reaching the ground state $\ket{1_g}$ (represented by the solid blue line) and the state $\ket{W_1}$ (indicated by the dashed red line) for a specific velocity $\upsilon$. In contrast, Fig.~\ref{fig_2}(c) displays the probabilities at the end of the pulse, with the matter subsystem initially prepared in the ground state.\\
\begin{figure}[t!]
\begin{center}
\includegraphics[width=0.95 \textwidth]{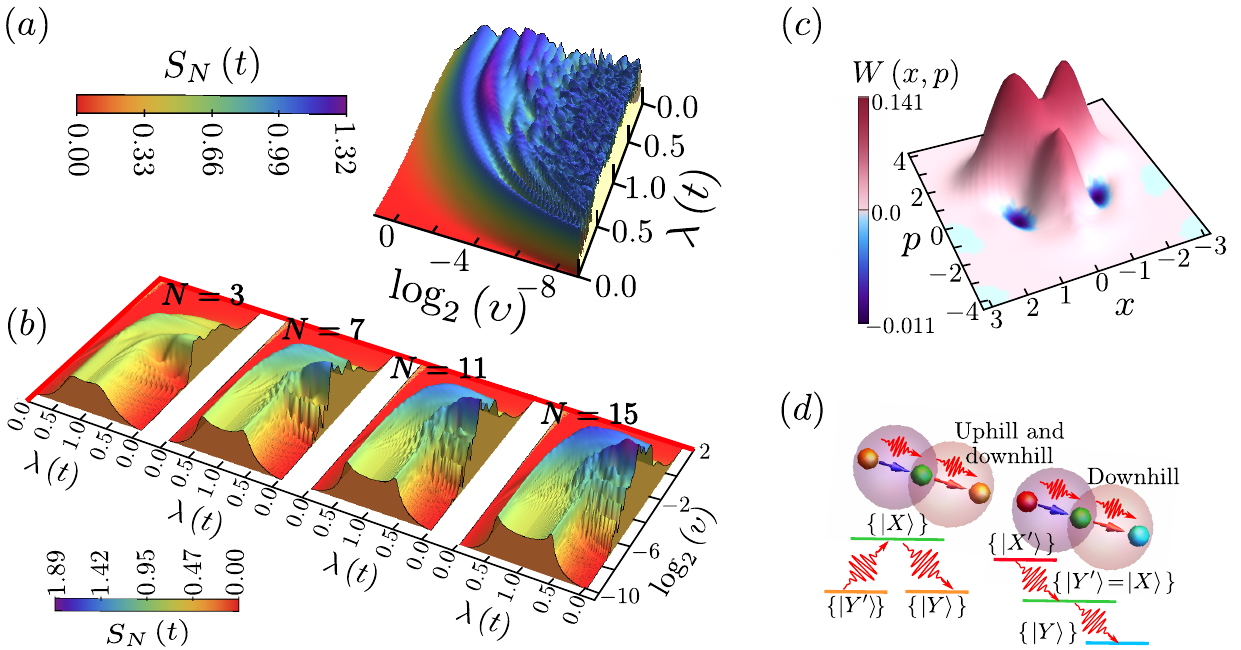}
\caption{\label{fig_3}{\bf (a)} Vibronic (i.e. electronic-vibrational) entanglement, as measured by the von Neumann entropy $S_N$, for $N=3$ system during an up-down pulse $\lambda(t)$, starting in the higher probability state $W_2$  as in Fig.~\ref{fig_2}(b). Shown as a function of the logarithm of the inverse pulse duration (i.e. ramping velocity) $\upsilon$. {\bf (b)} Vibronic (i.e. electronic-vibrational) entanglement $S_N$ for increasing $N$, starting in the initial $t=0$ ground state (i.e. $|\frac{3}{2},-\frac{3}{2}\rangle$ for $N=3$). In {\bf (a)} and {\bf (b)}, the vibronic entanglement (i.e. quantum coherence) is largest where shading is darkest. {\bf (c)} Vibrational subsystem Wigner function ($N=3$) for the final state after the up-down pulse, $\lambda(t)$, at intermediate $\log_2\pas{\upsilon} = -1.86$, confirming the non-classicality of the vibrational system's post-pulse state.  {\bf (d)} Schematic of possible energy transport across the system (Fig.~\ref{fig_1}(a)): during each step, the excitonic energy is interchanged with quantized vibrations, the total wavefunction is time-evolving, mutual entanglement between electronic and vibrational quanta develops, but the full system's state remains pure.}
 \end{center}
 \end{figure}
\subsection*{$N$-Body Entanglement}
In Figs.~\ref{fig_2}-\ref{fig_3}, we now show how pulses in the electronic-vibrational coupling $\lambda(t)$ enhance the generation, spreading, and eventual channeling of multi-exciton states across an $N$-component system (e.g. Fig.~\ref{fig_1}) and that the cause lies in the $N$-body vibronic entanglement that the pulse generates and manipulates in time (i.e. time-dependent quantum mechanical coherence between the electronic and vibrational subsystems, Fig.~\ref{fig_3}(b)). Figure~\ref{fig_2}(a) considers initial state given by Eq.~\eqref{gs_state}, which is the electronic part of the $t=0$ `ground' eigenstate of Eq.~\eqref{hdic} ($\lambda(0)=0$) comprising the lower excitonic states $\{|Y_\Gamma\rangle\}$ in any three near-resonant nanostructure components: for example, the three near-resonant dimers in Fig.~\ref{fig_1}(a). Additionally, after a single up-down pulse of moderate duration (i.e. intermediate $\upsilon$), the final state is dominated by $W_2$-states in which two excitations are shared among the three higher energy nanostructure components, and the entanglement spreads these maximally (Fig.~\ref{fig_2}(a) for $W_2$.\\
\\
This means that if quantum coherence within one of the three nanostructure components is lost, the state of the remaining two nanostructure components remains entangled. This dominant $W_2$ generation is due to the high vibronic entanglement (i.e. many-body quantum coherence as measured by the  von Neumann entropy $S_N$, Fig.~\ref{fig_3}(a)(b)) that is generated between the excitonic and vibrational subsystems by intermediate duration pulses $\lambda(t)$. For the electronic subsystem, the von Neumann entropy is defined as follows:
\begin{equation} \label{eqentro}
S_N (t)=-\tr{\rho_{e}\pap{t}\log \pas{\rho_e \pap{t}}}\;,\quad \text{where}\;\rho_e\pap{t}=\mathrm{tr}_v\pac{\ket{\psi\pap{t}}\bra{\psi \pap{t}}}
\end{equation}
In this context, we consider a bipartition where subsystem $v$ corresponds to the vibrational mode, and subsystem $e$ encompasses the molecular excitonic component, within a total pure state $\ket{\psi\pap{t}}$. In such a pure state, the entropy of subsystem $e$ is inherently equal to the entropy of its complementary subsystem $v$. This quantity, denoted as $S_N$, serves as a measure of the entanglement existing between these two subsystems.\\
\\
Since we are dealing with a closed system, characterized by a pure global quantum state that undergoes unitary evolution, any increase in $S_N$ within each subsystem signifies an exchange of information between the vibrational and matter constituents during the entire cycle. This observation lends itself to a more direct thermodynamic interpretation of the memory effects observed throughout the cycle. Figure~\ref{fig_3}(b) shows that the strength of this many-body quantum coherence and the range of pulse durations (i.e. inverse $\upsilon$) over which it exists, both {\em increase} with $N$. Unlike Fig.~\ref{fig_2}(a), the probability of generating two excited exciton states in three independent nanostructure components would be lower and the state would not be entangled. Our calculations therefore predict that an intermediate duration pulse $\lambda(t)$ will generate $N$-body vibronic entanglement that efficiently spreads multiple excited excitons within the system, and that this effect will become even stronger as $N$ increases.\\
\\
Figures~\ref{fig_2}(b), (c) and~\ref{fig_3}(a) illustrate the influence of a pulse $\lambda(t)$ on a system initially in the $W_2$ state, as generated previously. Notably, with a relatively high probability (approximately $\sim0.5$), the application of a moderate-duration pulse, characterized by intermediate $\upsilon$, facilitates the transfer of both excitons across one nanostructure component to another. This transfer culminates in the state $|Y_1,Y_2,Y_3\rangle$. This process of transferring excitations can be illustrated using a simple model depicting the movement and transport of excitons: a single dimer ($N=1$) starting in state $|X\rangle$, would produce a final state $|Y\rangle$ with probability $\sim0.5$, in line with simple arguments based on Rabi oscillations in a two-level system and Ref.~\cite{OReilly2014_5}: hence for three independent ($N=1$) dimers in the same two-excitation initial state (e.g. $|X_1\rangle |X_2\rangle |Y_3\rangle$), the corresponding probability of ending in $|Y_1\rangle |Y_2\rangle |Y_3\rangle$ is $\sim (0.5)^3=0.125$. Hence for the LHC-II, this would mean that the pulse $\lambda(t)$ has manipulated the $N$-body vibronic entanglement in order to more  efficiently transfer the excitons, and hence energy, to chlorophylls closest to the exit points across the trimer (i.e. to $\{|Y_\Gamma\rangle\}$).\\
\\
Irrespective of whether biophysical systems naturally use these features or not, these features could be exploited in future nanostructure device designs for energy and information processing~\cite{Rey2007,Hardal_CRP2015,NiedenzuPRE2015,Marquadt2011, Reslen2005epl, Acevedo2015NJP, AcevedoPRA2015,Acevedo2014PRL} -- in particular, using structures such as LHC-II given their natural abundance~\cite{Trimer,Drop,Lambrev}. This suggests that by changing the choice of dimer in Fig.~\ref{fig_1}(a), and hence the definition of lower and upper state and thus on which chromophores the excitations primarily lie, multiple highly-entangled excitations can be generated by individual pulses, transferred quantum mechanically between chromophore pairs across the trimer, and can hence reach the chromophores closest to exit sites, e.g. `uphill and downhill' $\{|Z\rangle\}\rightarrow\{|X\rangle\}\rightarrow\{|Y\rangle\}$ (Fig.~\ref{fig_3}(d)), and `downhill' $\{|X'\rangle\}\rightarrow\{|Y'\rangle\equiv |X\rangle\}\rightarrow\{|Y\rangle\}$, or any combination of these. This would occur without the need for energy relaxation processes to drive the direction of energy flow, hence the total wavefunction remains in a pure many-body quantum state --  and it does not require strong electronic-vibrational coupling.\\
\\
Though we take $\lambda(t)$ to be a piecewise linear up-down ramping of duration $\upsilon^{-1}$ (i.e. the inverse ramping velocity), similar results will occur for any other up-down functional form since the strong electronic-vibrational (i.e. vibronic) entanglement at intermediate $\upsilon$ has its roots in path interference caused by two crossings of the quantum critical point $\lambda_c=0.5$ \cite{AcevedoPRA2015, Acevedo2015NJP, GomezEnt2016}. Hence the details of the path are relatively unimportant. In particular, the shaded forms in  Figs.~\ref{fig_2}(a), 3(a), 3(b) can be understood by averaging over the quantum oscillations generated by the up-down path interference in a simple two-level Landau-Zener-Stuckelberg picture: the average probability that the system ends up in the excited state manifold is $P^e=2P(1-P)$, where $P={\rm exp}(-2\pi\Delta^2/4\upsilon)$ and $\Delta$ is the minimum effective two-level energy gap during the pulse. Approximating $\Delta\sim \lambda_c=0.5$, this predicts that $P^e$ increases monotonically as $\upsilon$ increases from the adiabatic regime, before falling off as ${\rm log}_2 (\upsilon)\rightarrow 0$, exactly as seen in Fig.~\ref{fig_2}(a),\ref{fig_3}(a),(b).\\
\\
Another method to analyze the indicators of non-classical behavior within the vibronic subsystem involves the Wigner quasi-probability distribution. The overall state of the mode is succinctly represented through its Wigner quasi-probability distribution~\cite{Scully1997},
\begin{equation}
W\pap{z,\rho_v}=\sum_{m=0}^{\infty}\pap{-1}^{m}\bra{m} \mathcal{D}^{\dagger}\pap{z}\rho_v\mathcal{D}\pap{z}\ket{m},
\end{equation}
In this context, the conventional displacement operator is defined as $\mathcal{D}(z) = \exp\left[z a^{\dagger} - z^*a\right]$, where $a^{\dagger}$ ($a$) represents the creation (annihilation) operator of the vibrionic mode, and $z$ is a complex number. Figures~\ref{fig_3}(a),(b),(c) confirm the intrinsic non-classical nature of these processes generated by the pulse $\lambda(t)$. Figure~\ref{fig_3}(b) indicates that as $N$ increases, the final states should be dominated by increasingly higher entangled equivalents of $W$ states that entangle successively higher numbers of excited exciton states ($\{|X\rangle\}$) which can then be manipulated by subsequent pulses to efficiently channel energy toward $\{|Y\rangle\}$ states and hence energy exit points. Given the naturally-occurring availability of  $N>3$ aggregates \cite{Trimer,Drop,Lambrev}, this enhancement with increasing $N$ may inspire new device designs. The lower bound $\upsilon_{\mathrm{min}}$ in Fig.~\ref{fig_3}(b) does not depend on the maximum value of $\lambda(t)$ reached. Instead, the scaling $\upsilon_{\mathrm{min}} \propto N^{-1}$ comes from a relation for the minimal energy gap at the critical threshold~\cite{AcevedoPRA2015}. The upper bound $\upsilon_{\mathrm{max}}$ does not depend on system size, and is instead dictated by the ${\rm log}_2 (v)\rightarrow 0$ drop-off of $P^e$. Figure~\ref{fig_3}(c) illustrates the purely quantum features that are present in the Wigner functions of the vibrational and electronic subsystems (e.g. negative values).\\
\\
Though our results consider ramping up to the modest value of $\lambda(t)\approx 1$ and back, similar results occur for smaller maximum values and hence weaker pulses as long as maximum $\lambda(t)\geq 0.5$. For maximum $\lambda(t)< 0.5$, the system does not feel the quantum critical point and hence there is negligible entanglement, in line with empirical observations that quantum coherence effects are primarily found beyond the weak perturbative driving field regime.
\subsection*{Robustness against decoherence/losses}
\begin{figure*}[t!]
\begin{center}
  \includegraphics[width=0.9\textwidth]{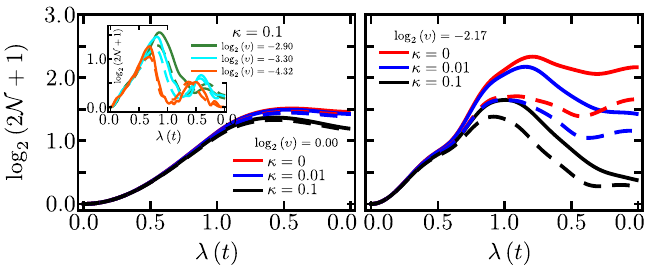}
  \caption{\label{fig_4} Evidence of robustness against decoherence/losses of the many-body electronic-vibrational entanglement (i.e. quantum coherence) as measured by quantum logarithmic negativity (see text). Results are shown for two representative intermediate up-down pulse durations (i.e. different $v$ in left and right panels). Results shown for $N=5$ (dashed lines) and $N=11$ (solid lines), and various values of decoherence $\kappa$. As decoherence increases, the differences between the curves for different $N$ tend to become smaller: this hints at a possible universality in robustness with increasing $\kappa$. Inset: largest $\kappa$ value and different $\upsilon$. We find that increasing temperature (and hence $\langle n\rangle$) shows broadly similar tendencies to the results shown for increasing $\kappa$.} 
  \end{center}
 \end{figure*}
The presence of decoherence/losses to and from the environment, does not change our main conclusions, as illustrated in Fig.~\ref{fig_4}. Since $S_N$ is no longer a good entanglement measure in an open system, we use the closely related quantity quantum negativity~\cite{Vidalneg} $\mathcal{N}\pap{\rho}=\frac{1}{2}(\paV{\rho^{\Gamma_q}}_{1}-1)$ where $\rho^{\Gamma_q}$ is the partial transpose of $\rho$ with respect to the matter subsystem, and $\paV{A}_{1}\equiv\mathrm{tr}\pac{\sqrt{A^{\dagger}A}}$ is the trace norm.
It gives essentially the same results as $S_N$ for small decoherence since both measures represent essentially the same information~\cite{Johnson}. The electronic-vibrational density matrix $\rho (t)$ evolves as~\cite{breuer2002theory}:
\begin{equation}\label{eqME}
{\dot\rho} = - i \pas{H,\rho} +2\kappa \pap{\bar{n}+1} \mathcal{L} \pap{\rho;a}+2\kappa \bar{n} \mathcal{L} \pap{\rho;a^{\dagger}},
\end{equation}
where $\kappa$ is the damping rate and $\bar{n}$ is the thermal mean photon number. The Lindblad superoperator $\mathcal{L}$ is defined as $A \rho A^{\dagger}-\frac{1}{2}\pac{A^{\dagger}A,\rho}$ ($\pac{\bullet,\bullet}$ is the anti-commutator). Not only do our main results survive well with increasing decoherence $\kappa$, the strength and robustness of the many-body coherence both increase with $N$. (For very large $N$, other system-level decoherence mechanisms may set in). We clarify that in this analysis, we have neglected the energy level fluctuations. A more realistic approach for photosynthetic systems would include (see, for example, Ref.~\cite{Sangwoo}). We have also simulated different temperatures by varying the average number of quantized vibrations $\bar{n}$, choosing values typical of the low temperatures in most experimental realizations, and including a thermal distribution in the initial density matrix~\cite{AcevedoPRA2015}. Again we find that our main conclusions are qualitatively unchanged.\\
\\
\subsection*{Further observations} 

In addition to the specific results shown above, it is worth noting the following general points:\\

\noindent (1) Our focus in this paper is on the question of collective coherence in any setting where there is some pulsed perturbation of the system, in order to explore better the idea of how excitations (e.g. excitons) may `ride a wave' of coherence. However the external profile ${\vec E}(t)$ and hence $\lambda(t)$ can be general. It need not be a pulse. Also the application of Eq.~\eqref{hdic} could be to transport experiments as well as optical experiments, or any combination of these.\\ 

\noindent (2)  There are two ways in which Eq.~\eqref{hdic} can be applied: (i) We assume that incident light has already created excitons in our system, hence the initial ground state in Eq.~\eqref{hdic} is one in which each of the $N$ nanostructure components is in the lower excited state, i.e. $\ket{J,J_z=-N/2}$. An initial excited state in Eq.~\eqref{hdic} is one in which one or more of the $N$ nanostructure components is in the upper excited state, i.e. $ \ket{J,J_z>-N/2}$. Our model then calculates what additional many-body coherence is generated by a pulse. (ii) We can alternatively assume that the system starts in a true ground state with no excitations. The same analysis follows. It is case (i) that would be relevant for the particular setting of LHC-II, since it is the energy separation of each dimer's two  hybridized excitonic states $|Y\rangle$ and $|X\rangle$ that is quasi-resonant with a vibrational mode.\\ 

\noindent (3)  The pulse $\lambda(t)$ is a consequence of the internal polarization ${\vec P}(t)$ due to an external pulse of light ${\vec E}(t)$. It is well known that Maxwell's equations give a nonlinear, exact equation that relates ${\vec P}(t)$ to ${\vec E}(t)$, and as a result of the mathematical form (see above) a pulse in ${\vec E}(t)$ will generate a pulse in ${\vec P}(t)$. This justifies the appearance of the pulse in our time-dependent Hamiltonian. \\ %This can also be established starting from a full quantum picture (see Supplementary Material).\\

\noindent (4)  The effect of finite temperature is treated when discussing the losses/decoherence, by including finite numbers of vibrational quanta consistent with temperature distributions. We find that these do not change our main conclusions for reasonably low temperatures. Even for higher temperatures, the effects that we discuss have not disappeared.\\

\noindent (5)  Several papers (e.g. Ref.~\cite{Jianshu_SciEDv2020,Donatas2022,Lambert2013}) make the point that a deeper understanding of the temporal quantum evolution in systems of general size $N$ -- as this study tries to offer -- may shed light on the early-time kinetics in real-world open systems, since this timescale is too short to couple in many of the complex degrees of freedom that will naturally exist in a hot, wet environment. 
\section*{SUMMARY AND PERSPECTIVES}
We have demonstrated that pulse-driven perturbations give rise to novel dynamical phenomena linked to many-body ($N\geq 3$) entanglement, specifically quantum coherence. These phenomena can be harnessed for energy harvesting, manipulation, and the design of quantum information devices, all without the necessity of strong electronic-vibrational coupling. By contrast, existing theoretical investigations tend to concentrate on scenarios where $N\rightarrow 1$ and/or use perturbation theory and/or  average over memory effects, and hence have not yet uncovered these intriguing new physical phenomena reported here -- despite potentially offering a more chemically detailed picture of the system itself.

\section*{Data availability}
The data generated during the current study are available from the corresponding author on reasonable request. All the data and results during the current study are generated by the mathematical model and code. 
\section*{Code availability}
The codes for this study are available from the corresponding author on reasonable request. Figures created by the authors using Mathematica version 13 \url{https://www.wolfram.com/mathematica/} and also Microsoft Powerpoint version 16 \url{https://www.microsoft.com/en-us/microsoft-365}
\section*{Acknowledgements}
F.J.G-R acknowledges financial support from European Commission FET-Open project AVaQus GA 899561. This research was supported by Spanish MCIN with funding from European Union Next Generation EU (PRTRC17.I1) and Consejeria de Educacion from JCyL through QCAYLE project, as well as MCIN projects PID2020-113406GB-I00 and RED2022-134301-T.  F.J.R. and L.Q. are thankful for financial support from Facultad de Ciencias-UniAndes projects INV-2021-128-2292 and INV-2023-162-2833. N.F.J. is supported by U.S. Air Force Office of Scientific Research awards FA9550-20-1-0382 and FA9550-20-1-0383. The views and conclusions contained herein are solely those of the authors and do not represent official policies or endorsements by any of the entities named in this paper.
\section*{Author contributions}
F. J. Gómez-Ruiz developed numerical simulations and prepared the figures. All authors contributed to the analysis of the results and the writing of the manuscript. N. F. Johnson and L. Quiroga initiated and guided the project.	
\section*{Competing interests}
The authors declare no competing interests.	
\bibliography{mybib_SReport}{}

\begin{thebibliography}{74}%
\makeatletter
\providecommand \@ifxundefined [1]{%
 \@ifx{#1\undefined}
}%
\providecommand \@ifnum [1]{%
 \ifnum #1\expandafter \@firstoftwo
 \else \expandafter \@secondoftwo
 \fi
}%
\providecommand \@ifx [1]{%
 \ifx #1\expandafter \@firstoftwo
 \else \expandafter \@secondoftwo
 \fi
}%
\providecommand \natexlab [1]{#1}%
\providecommand \enquote  [1]{``#1''}%
\providecommand \bibnamefont  [1]{#1}%
\providecommand \bibfnamefont [1]{#1}%
\providecommand \citenamefont [1]{#1}%
\providecommand \href@noop [0]{\@secondoftwo}%
\providecommand \href [0]{\begingroup \@sanitize@url \@href}%
\providecommand \@href[1]{\@@startlink{#1}\@@href}%
\providecommand \@@href[1]{\endgroup#1\@@endlink}%
\providecommand \@sanitize@url [0]{\catcode `\\12\catcode `\$12\catcode
  `\&12\catcode `\#12\catcode `\^12\catcode `\_12\catcode `\%12\relax}%
\providecommand \@@startlink[1]{}%
\providecommand \@@endlink[0]{}%
\providecommand \url  [0]{\begingroup\@sanitize@url \@url }%
\providecommand \@url [1]{\endgroup\@href {#1}{\urlprefix }}%
\providecommand \urlprefix  [0]{URL }%
\providecommand \Eprint [0]{\href }%
\providecommand \doibase [0]{http://dx.doi.org/}%
\providecommand \selectlanguage [0]{\@gobble}%
\providecommand \bibinfo  [0]{\@secondoftwo}%
\providecommand \bibfield  [0]{\@secondoftwo}%
\providecommand \translation [1]{[#1]}%
\providecommand \BibitemOpen [0]{}%
\providecommand \bibitemStop [0]{}%
\providecommand \bibitemNoStop [0]{.\EOS\space}%
\providecommand \EOS [0]{\spacefactor3000\relax}%
\providecommand \BibitemShut  [1]{\csname bibitem#1\endcsname}%
\let\auto@bib@innerbib\@empty
%</preamble>
\bibitem [{\citenamefont {Nishida}\ \emph {et~al.}(2022)\citenamefont
  {Nishida}, \citenamefont {Johnson}, \citenamefont {Chang}, \citenamefont
  {Wharton}, \citenamefont {D{\"o}nges}, \citenamefont {Khatib},\ and\
  \citenamefont {Raschke}}]{Nishida_NatCom_22}%
  \BibitemOpen
  \bibfield  {author} {\bibinfo {author} {\bibfnamefont {J.}~\bibnamefont
  {Nishida}}, \bibinfo {author} {\bibfnamefont {S.~C.}\ \bibnamefont
  {Johnson}}, \bibinfo {author} {\bibfnamefont {P.~T.~S.}\ \bibnamefont
  {Chang}}, \bibinfo {author} {\bibfnamefont {D.~M.}\ \bibnamefont {Wharton}},
  \bibinfo {author} {\bibfnamefont {S.~A.}\ \bibnamefont {D{\"o}nges}},
  \bibinfo {author} {\bibfnamefont {O.}~\bibnamefont {Khatib}}, \ and\ \bibinfo
  {author} {\bibfnamefont {M.~B.}\ \bibnamefont {Raschke}},\ }\href {\doibase
  10.1038/s41467-022-28224-9} {\bibfield  {journal} {\bibinfo  {journal}
  {Nature Communications}\ }\textbf {\bibinfo {volume} {13}},\ \bibinfo {pages}
  {1083} (\bibinfo {year} {2022})}\BibitemShut {NoStop}%
\bibitem [{\citenamefont {Cao}\ \emph {et~al.}(2020)\citenamefont {Cao},
  \citenamefont {Cogdell}, \citenamefont {Coker}, \citenamefont {Duan},
  \citenamefont {Hauer}, \citenamefont {Kleinekathöfer}, \citenamefont
  {Jansen}, \citenamefont {Mančal}, \citenamefont {Miller}, \citenamefont
  {Ogilvie}, \citenamefont {Prokhorenko}, \citenamefont {Renger}, \citenamefont
  {Tan}, \citenamefont {Tempelaar}, \citenamefont {Thorwart}, \citenamefont
  {Thyrhaug}, \citenamefont {Westenhoff},\ and\ \citenamefont
  {Zigmantas}}]{Jianshu_SciEDv2020}%
  \BibitemOpen
  \bibfield  {author} {\bibinfo {author} {\bibfnamefont {J.}~\bibnamefont
  {Cao}}, \bibinfo {author} {\bibfnamefont {R.~J.}\ \bibnamefont {Cogdell}},
  \bibinfo {author} {\bibfnamefont {D.~F.}\ \bibnamefont {Coker}}, \bibinfo
  {author} {\bibfnamefont {H.-G.}\ \bibnamefont {Duan}}, \bibinfo {author}
  {\bibfnamefont {J.}~\bibnamefont {Hauer}}, \bibinfo {author} {\bibfnamefont
  {U.}~\bibnamefont {Kleinekathöfer}}, \bibinfo {author} {\bibfnamefont
  {T.~L.~C.}\ \bibnamefont {Jansen}}, \bibinfo {author} {\bibfnamefont
  {T.}~\bibnamefont {Mančal}}, \bibinfo {author} {\bibfnamefont {R.~J.~D.}\
  \bibnamefont {Miller}}, \bibinfo {author} {\bibfnamefont {J.~P.}\
  \bibnamefont {Ogilvie}}, \bibinfo {author} {\bibfnamefont {V.~I.}\
  \bibnamefont {Prokhorenko}}, \bibinfo {author} {\bibfnamefont
  {T.}~\bibnamefont {Renger}}, \bibinfo {author} {\bibfnamefont {H.-S.}\
  \bibnamefont {Tan}}, \bibinfo {author} {\bibfnamefont {R.}~\bibnamefont
  {Tempelaar}}, \bibinfo {author} {\bibfnamefont {M.}~\bibnamefont {Thorwart}},
  \bibinfo {author} {\bibfnamefont {E.}~\bibnamefont {Thyrhaug}}, \bibinfo
  {author} {\bibfnamefont {S.}~\bibnamefont {Westenhoff}}, \ and\ \bibinfo
  {author} {\bibfnamefont {D.}~\bibnamefont {Zigmantas}},\ }\href {\doibase
  10.1126/sciadv.aaz4888} {\bibfield  {journal} {\bibinfo  {journal} {Science
  Advances}\ }\textbf {\bibinfo {volume} {6}},\ \bibinfo {pages} {eaaz4888}
  (\bibinfo {year} {2020})}\BibitemShut {NoStop}%
\bibitem [{\citenamefont {Lambert}\ \emph {et~al.}(2013)\citenamefont
  {Lambert}, \citenamefont {Chen}, \citenamefont {Cheng}, \citenamefont {Li},
  \citenamefont {Chen},\ and\ \citenamefont {Nori}}]{Lambert2013}%
  \BibitemOpen
  \bibfield  {author} {\bibinfo {author} {\bibfnamefont {N.}~\bibnamefont
  {Lambert}}, \bibinfo {author} {\bibfnamefont {Y.-N.}\ \bibnamefont {Chen}},
  \bibinfo {author} {\bibfnamefont {Y.-C.}\ \bibnamefont {Cheng}}, \bibinfo
  {author} {\bibfnamefont {C.-M.}\ \bibnamefont {Li}}, \bibinfo {author}
  {\bibfnamefont {G.-Y.}\ \bibnamefont {Chen}}, \ and\ \bibinfo {author}
  {\bibfnamefont {F.}~\bibnamefont {Nori}},\ }\href {\doibase
  10.1038/nphys2474} {\bibfield  {journal} {\bibinfo  {journal} {Nature
  Physics}\ }\textbf {\bibinfo {volume} {9}},\ \bibinfo {pages} {10} (\bibinfo
  {year} {2013})}\BibitemShut {NoStop}%
\bibitem [{\citenamefont {Scholes}\ \emph {et~al.}(2017)\citenamefont
  {Scholes}, \citenamefont {Fleming}, \citenamefont {Chen}, \citenamefont
  {Aspuru-Guzik}, \citenamefont {Buchleitner}, \citenamefont {Coker},
  \citenamefont {Engel}, \citenamefont {van Grondelle}, \citenamefont
  {Ishizaki}, \citenamefont {Jonas}, \citenamefont {Lundeen}, \citenamefont
  {McCusker}, \citenamefont {Mukamel}, \citenamefont {Ogilvie}, \citenamefont
  {Olaya-Castro}, \citenamefont {Ratner}, \citenamefont {Spano}, \citenamefont
  {Whaley},\ and\ \citenamefont {Zhu}}]{Scholes_1}%
  \BibitemOpen
  \bibfield  {author} {\bibinfo {author} {\bibfnamefont {G.~D.}\ \bibnamefont
  {Scholes}}, \bibinfo {author} {\bibfnamefont {G.~R.}\ \bibnamefont
  {Fleming}}, \bibinfo {author} {\bibfnamefont {L.~X.}\ \bibnamefont {Chen}},
  \bibinfo {author} {\bibfnamefont {A.}~\bibnamefont {Aspuru-Guzik}}, \bibinfo
  {author} {\bibfnamefont {A.}~\bibnamefont {Buchleitner}}, \bibinfo {author}
  {\bibfnamefont {D.~F.}\ \bibnamefont {Coker}}, \bibinfo {author}
  {\bibfnamefont {G.~S.}\ \bibnamefont {Engel}}, \bibinfo {author}
  {\bibfnamefont {R.}~\bibnamefont {van Grondelle}}, \bibinfo {author}
  {\bibfnamefont {A.}~\bibnamefont {Ishizaki}}, \bibinfo {author}
  {\bibfnamefont {D.~M.}\ \bibnamefont {Jonas}}, \bibinfo {author}
  {\bibfnamefont {J.~S.}\ \bibnamefont {Lundeen}}, \bibinfo {author}
  {\bibfnamefont {J.~K.}\ \bibnamefont {McCusker}}, \bibinfo {author}
  {\bibfnamefont {S.}~\bibnamefont {Mukamel}}, \bibinfo {author} {\bibfnamefont
  {J.~P.}\ \bibnamefont {Ogilvie}}, \bibinfo {author} {\bibfnamefont
  {A.}~\bibnamefont {Olaya-Castro}}, \bibinfo {author} {\bibfnamefont {M.~A.}\
  \bibnamefont {Ratner}}, \bibinfo {author} {\bibfnamefont {F.~C.}\
  \bibnamefont {Spano}}, \bibinfo {author} {\bibfnamefont {K.~B.}\ \bibnamefont
  {Whaley}}, \ and\ \bibinfo {author} {\bibfnamefont {X.}~\bibnamefont {Zhu}},\
  }\href {\doibase 10.1038/nature21425} {\bibfield  {journal} {\bibinfo
  {journal} {Nature}\ }\textbf {\bibinfo {volume} {543}},\ \bibinfo {pages}
  {647} (\bibinfo {year} {2017})}\BibitemShut {NoStop}%
\bibitem [{\citenamefont {O'Reilly}\ and\ \citenamefont
  {Olaya-Castro}(2014)}]{OReilly2014_5}%
  \BibitemOpen
  \bibfield  {author} {\bibinfo {author} {\bibfnamefont {E.~J.}\ \bibnamefont
  {O'Reilly}}\ and\ \bibinfo {author} {\bibfnamefont {A.}~\bibnamefont
  {Olaya-Castro}},\ }\href {\doibase 10.1038/ncomms4012} {\bibfield  {journal}
  {\bibinfo  {journal} {Nature Communications}\ }\textbf {\bibinfo {volume}
  {5}} (\bibinfo {year} {2014}),\ 10.1038/ncomms4012}\BibitemShut {NoStop}%
\bibitem [{\citenamefont {Plenio}\ \emph {et~al.}(2013)\citenamefont {Plenio},
  \citenamefont {Almeida},\ and\ \citenamefont {Huelga}}]{Plenio_23}%
  \BibitemOpen
  \bibfield  {author} {\bibinfo {author} {\bibfnamefont {M.~B.}\ \bibnamefont
  {Plenio}}, \bibinfo {author} {\bibfnamefont {J.}~\bibnamefont {Almeida}}, \
  and\ \bibinfo {author} {\bibfnamefont {S.~F.}\ \bibnamefont {Huelga}},\
  }\href {\doibase 10.1063/1.4846275} {\bibfield  {journal} {\bibinfo
  {journal} {J. Chem. Phys.}\ }\textbf {\bibinfo {volume} {139}},\ \bibinfo
  {pages} {235102} (\bibinfo {year} {2013})}\BibitemShut {NoStop}%
\bibitem [{\citenamefont {Zigmantas}\ \emph {et~al.}(2022)\citenamefont
  {Zigmantas}, \citenamefont {Pol\'ivka}, \citenamefont {Persson},\ and\
  \citenamefont {Sundstr\"om}}]{Donatas2022}%
  \BibitemOpen
  \bibfield  {author} {\bibinfo {author} {\bibfnamefont {D.}~\bibnamefont
  {Zigmantas}}, \bibinfo {author} {\bibfnamefont {T.}~\bibnamefont
  {Pol\'ivka}}, \bibinfo {author} {\bibfnamefont {P.}~\bibnamefont {Persson}},
  \ and\ \bibinfo {author} {\bibfnamefont {V.}~\bibnamefont {Sundstr\"om}},\
  }\href {\doibase 10.1063/5.0092864} {\bibfield  {journal} {\bibinfo
  {journal} {Chemical Physics Reviews}\ }\textbf {\bibinfo {volume} {3}}
  (\bibinfo {year} {2022}),\ 10.1063/5.0092864},\ \bibinfo {note}
  {041303}\BibitemShut {NoStop}%
\bibitem [{\citenamefont {Olaya-Castro}\ \emph {et~al.}(2008)\citenamefont
  {Olaya-Castro}, \citenamefont {Lee}, \citenamefont {Olsen},\ and\
  \citenamefont {Johnson}}]{Olaya_PRB08}%
  \BibitemOpen
  \bibfield  {author} {\bibinfo {author} {\bibfnamefont {A.}~\bibnamefont
  {Olaya-Castro}}, \bibinfo {author} {\bibfnamefont {C.~F.}\ \bibnamefont
  {Lee}}, \bibinfo {author} {\bibfnamefont {F.~F.}\ \bibnamefont {Olsen}}, \
  and\ \bibinfo {author} {\bibfnamefont {N.~F.}\ \bibnamefont {Johnson}},\
  }\href {\doibase 10.1103/PhysRevB.78.085115} {\bibfield  {journal} {\bibinfo
  {journal} {Phys. Rev. B}\ }\textbf {\bibinfo {volume} {78}},\ \bibinfo
  {pages} {085115} (\bibinfo {year} {2008})}\BibitemShut {NoStop}%
\bibitem [{\citenamefont {Thilagam}(2015)}]{Thila}%
  \BibitemOpen
  \bibfield  {author} {\bibinfo {author} {\bibfnamefont {A.}~\bibnamefont
  {Thilagam}},\ }\href {\doibase 10.1007/s10910-014-0442-x} {\bibfield
  {journal} {\bibinfo  {journal} {J. Math. Chem.}\ }\textbf {\bibinfo {volume}
  {53}},\ \bibinfo {pages} {466} (\bibinfo {year} {2015})}\BibitemShut
  {NoStop}%
\bibitem [{\citenamefont {Fujihashi}\ \emph {et~al.}(2015)\citenamefont
  {Fujihashi}, \citenamefont {Fleming},\ and\ \citenamefont
  {Ishizaki}}]{Fujihashi_3}%
  \BibitemOpen
  \bibfield  {author} {\bibinfo {author} {\bibfnamefont {Y.}~\bibnamefont
  {Fujihashi}}, \bibinfo {author} {\bibfnamefont {G.~R.}\ \bibnamefont
  {Fleming}}, \ and\ \bibinfo {author} {\bibfnamefont {A.}~\bibnamefont
  {Ishizaki}},\ }\href {\doibase 10.1063/1.4914302} {\bibfield  {journal}
  {\bibinfo  {journal} {J. Chem. Phys.}\ }\textbf {\bibinfo {volume} {142}},\
  \bibinfo {pages} {212403} (\bibinfo {year} {2015})}\BibitemShut {NoStop}%
\bibitem [{\citenamefont {Chin}\ \emph {et~al.}(2012)\citenamefont {Chin},
  \citenamefont {Prior}, \citenamefont {Rosenbach}, \citenamefont
  {Caycedo-Soler}, \citenamefont {Huelga},\ and\ \citenamefont
  {Plenio}}]{Chin_4}%
  \BibitemOpen
  \bibfield  {author} {\bibinfo {author} {\bibfnamefont {A.~W.}\ \bibnamefont
  {Chin}}, \bibinfo {author} {\bibfnamefont {J.}~\bibnamefont {Prior}},
  \bibinfo {author} {\bibfnamefont {R.}~\bibnamefont {Rosenbach}}, \bibinfo
  {author} {\bibfnamefont {F.}~\bibnamefont {Caycedo-Soler}}, \bibinfo {author}
  {\bibfnamefont {S.~F.}\ \bibnamefont {Huelga}}, \ and\ \bibinfo {author}
  {\bibfnamefont {M.~B.}\ \bibnamefont {Plenio}},\ }\href {\doibase
  10.1038/nphys2515} {\bibfield  {journal} {\bibinfo  {journal} {Nat. Phys.}\
  }\textbf {\bibinfo {volume} {9}},\ \bibinfo {pages} {113} (\bibinfo {year}
  {2012})}\BibitemShut {NoStop}%
\bibitem [{\citenamefont {Scholes}\ \emph {et~al.}(2011)\citenamefont
  {Scholes}, \citenamefont {Fleming}, \citenamefont {Olaya-Castro},\ and\
  \citenamefont {van Grondelle}}]{Scholes_11}%
  \BibitemOpen
  \bibfield  {author} {\bibinfo {author} {\bibfnamefont {G.~D.}\ \bibnamefont
  {Scholes}}, \bibinfo {author} {\bibfnamefont {G.~R.}\ \bibnamefont
  {Fleming}}, \bibinfo {author} {\bibfnamefont {A.}~\bibnamefont
  {Olaya-Castro}}, \ and\ \bibinfo {author} {\bibfnamefont {R.}~\bibnamefont
  {van Grondelle}},\ }\href {\doibase http://dx.doi.org/10.1038/nchem.1145}
  {\bibfield  {journal} {\bibinfo  {journal} {Nat. Chem.}\ }\textbf {\bibinfo
  {volume} {3}},\ \bibinfo {pages} {763} (\bibinfo {year} {2011})}\BibitemShut
  {NoStop}%
\bibitem [{\citenamefont {Engel}\ \emph {et~al.}(2007)\citenamefont {Engel},
  \citenamefont {Calhoun}, \citenamefont {Read}, \citenamefont {Ahn},
  \citenamefont {Mancal}, \citenamefont {Cheng}, \citenamefont {Blankenship},\
  and\ \citenamefont {Fleming}}]{Fleming_13}%
  \BibitemOpen
  \bibfield  {author} {\bibinfo {author} {\bibfnamefont {G.~S.}\ \bibnamefont
  {Engel}}, \bibinfo {author} {\bibfnamefont {T.~R.}\ \bibnamefont {Calhoun}},
  \bibinfo {author} {\bibfnamefont {E.~L.}\ \bibnamefont {Read}}, \bibinfo
  {author} {\bibfnamefont {T.}~\bibnamefont {Ahn}}, \bibinfo {author}
  {\bibfnamefont {T.}~\bibnamefont {Mancal}}, \bibinfo {author} {\bibfnamefont
  {Y.}~\bibnamefont {Cheng}}, \bibinfo {author} {\bibfnamefont {R.~E.}\
  \bibnamefont {Blankenship}}, \ and\ \bibinfo {author} {\bibfnamefont {G.~R.}\
  \bibnamefont {Fleming}},\ }\href {\doibase 10.1038/nature05678} {\bibfield
  {journal} {\bibinfo  {journal} {Nature}\ }\textbf {\bibinfo {volume} {446}},\
  \bibinfo {pages} {782} (\bibinfo {year} {2007})}\BibitemShut {NoStop}%
\bibitem [{\citenamefont {Collini}\ \emph {et~al.}(2010)\citenamefont
  {Collini}, \citenamefont {Wong}, \citenamefont {Wilk}, \citenamefont {Curmi},
  \citenamefont {Brumer},\ and\ \citenamefont {Scholes}}]{Collini_15}%
  \BibitemOpen
  \bibfield  {author} {\bibinfo {author} {\bibfnamefont {E.}~\bibnamefont
  {Collini}}, \bibinfo {author} {\bibfnamefont {C.~Y.}\ \bibnamefont {Wong}},
  \bibinfo {author} {\bibfnamefont {K.~E.}\ \bibnamefont {Wilk}}, \bibinfo
  {author} {\bibfnamefont {P.~M.~G.}\ \bibnamefont {Curmi}}, \bibinfo {author}
  {\bibfnamefont {P.}~\bibnamefont {Brumer}}, \ and\ \bibinfo {author}
  {\bibfnamefont {G.~D.}\ \bibnamefont {Scholes}},\ }\href {\doibase
  10.1038/nature08811} {\bibfield  {journal} {\bibinfo  {journal} {Nature}\
  }\textbf {\bibinfo {volume} {463}},\ \bibinfo {pages} {644} (\bibinfo {year}
  {2010})}\BibitemShut {NoStop}%
\bibitem [{\citenamefont {Rozzi}\ \emph {et~al.}(2013)\citenamefont {Rozzi},
  \citenamefont {Maria~Falke}, \citenamefont {Spallanzani}, \citenamefont
  {Rubio}, \citenamefont {Molinari}, \citenamefont {Brida}, \citenamefont
  {Maiuri}, \citenamefont {Cerullo}, \citenamefont {Schramm}, \citenamefont
  {Christoffers},\ and\ \citenamefont {Lienau}}]{Rozzi_17}%
  \BibitemOpen
  \bibfield  {author} {\bibinfo {author} {\bibfnamefont {C.~A.}\ \bibnamefont
  {Rozzi}}, \bibinfo {author} {\bibfnamefont {S.}~\bibnamefont {Maria~Falke}},
  \bibinfo {author} {\bibfnamefont {N.}~\bibnamefont {Spallanzani}}, \bibinfo
  {author} {\bibfnamefont {A.}~\bibnamefont {Rubio}}, \bibinfo {author}
  {\bibfnamefont {E.}~\bibnamefont {Molinari}}, \bibinfo {author}
  {\bibfnamefont {D.}~\bibnamefont {Brida}}, \bibinfo {author} {\bibfnamefont
  {M.}~\bibnamefont {Maiuri}}, \bibinfo {author} {\bibfnamefont
  {G.}~\bibnamefont {Cerullo}}, \bibinfo {author} {\bibfnamefont
  {H.}~\bibnamefont {Schramm}}, \bibinfo {author} {\bibfnamefont
  {J.}~\bibnamefont {Christoffers}}, \ and\ \bibinfo {author} {\bibfnamefont
  {C.}~\bibnamefont {Lienau}},\ }\href {\doibase 10.1038/ncomms2603} {\bibfield
   {journal} {\bibinfo  {journal} {Nat. Commun.}\ }\textbf {\bibinfo {volume}
  {4}},\ \bibinfo {pages} {1602} (\bibinfo {year} {2013})}\BibitemShut
  {NoStop}%
\bibitem [{\citenamefont {Tiwari}\ \emph {et~al.}(2013)\citenamefont {Tiwari},
  \citenamefont {Peters},\ and\ \citenamefont {Jonas}}]{Tiwari_22}%
  \BibitemOpen
  \bibfield  {author} {\bibinfo {author} {\bibfnamefont {V.}~\bibnamefont
  {Tiwari}}, \bibinfo {author} {\bibfnamefont {W.~K.}\ \bibnamefont {Peters}},
  \ and\ \bibinfo {author} {\bibfnamefont {D.~M.}\ \bibnamefont {Jonas}},\
  }\href {\doibase 10.1073/pnas.1211157110} {\bibfield  {journal} {\bibinfo
  {journal} {Proc. Nat. Acad. Sci. USA}\ }\textbf {\bibinfo {volume} {110}},\
  \bibinfo {pages} {1203} (\bibinfo {year} {2013})}\BibitemShut {NoStop}%
\bibitem [{\citenamefont {Schlau-Cohen}\ \emph {et~al.}(2012)\citenamefont
  {Schlau-Cohen}, \citenamefont {Ishizaki}, \citenamefont {Calhoun},
  \citenamefont {Ginsberg}, \citenamefont {Ballottari}, \citenamefont {Bassi},\
  and\ \citenamefont {Fleming}}]{Schlau_26}%
  \BibitemOpen
  \bibfield  {author} {\bibinfo {author} {\bibfnamefont {G.~S.}\ \bibnamefont
  {Schlau-Cohen}}, \bibinfo {author} {\bibfnamefont {A.}~\bibnamefont
  {Ishizaki}}, \bibinfo {author} {\bibfnamefont {T.~R.}\ \bibnamefont
  {Calhoun}}, \bibinfo {author} {\bibfnamefont {N.~S.}\ \bibnamefont
  {Ginsberg}}, \bibinfo {author} {\bibfnamefont {M.}~\bibnamefont
  {Ballottari}}, \bibinfo {author} {\bibfnamefont {R.}~\bibnamefont {Bassi}}, \
  and\ \bibinfo {author} {\bibfnamefont {G.~R.}\ \bibnamefont {Fleming}},\
  }\href {\doibase 10.1038/nchem.1303} {\bibfield  {journal} {\bibinfo
  {journal} {Nat. Chem.}\ }\textbf {\bibinfo {volume} {4}},\ \bibinfo {pages}
  {389} (\bibinfo {year} {2012})}\BibitemShut {NoStop}%
\bibitem [{\citenamefont {Br{\'e}das}\ \emph {et~al.}(2016)\citenamefont
  {Br{\'e}das}, \citenamefont {Sargent},\ and\ \citenamefont
  {Scholes}}]{Scholes_27}%
  \BibitemOpen
  \bibfield  {author} {\bibinfo {author} {\bibfnamefont {J.~L.}\ \bibnamefont
  {Br{\'e}das}}, \bibinfo {author} {\bibfnamefont {E.~H.}\ \bibnamefont
  {Sargent}}, \ and\ \bibinfo {author} {\bibfnamefont {G.~D.}\ \bibnamefont
  {Scholes}},\ }\href {\doibase 10.1038/nmat4767} {\bibfield  {journal}
  {\bibinfo  {journal} {Nat. Mater.}\ }\textbf {\bibinfo {volume} {16}},\
  \bibinfo {pages} {35} (\bibinfo {year} {2016})}\BibitemShut {NoStop}%
\bibitem [{\citenamefont {Scholes}\ and\ \citenamefont
  {Fleming}(2000)}]{Scholes_30}%
  \BibitemOpen
  \bibfield  {author} {\bibinfo {author} {\bibfnamefont {G.~D.}\ \bibnamefont
  {Scholes}}\ and\ \bibinfo {author} {\bibfnamefont {G.~R.}\ \bibnamefont
  {Fleming}},\ }\href {\doibase 10.1021/jp993435l} {\bibfield  {journal}
  {\bibinfo  {journal} {J. Phys. Chem. B}\ }\textbf {\bibinfo {volume} {104}},\
  \bibinfo {pages} {1854} (\bibinfo {year} {2000})}\BibitemShut {NoStop}%
\bibitem [{\citenamefont {Hainer}\ \emph {et~al.}(2021)\citenamefont {Hainer},
  \citenamefont {Alagna}, \citenamefont {Reddy~Marri}, \citenamefont {Penfold},
  \citenamefont {Gros}, \citenamefont {Haacke},\ and\ \citenamefont
  {Buckup}}]{Hainer2021}%
  \BibitemOpen
  \bibfield  {author} {\bibinfo {author} {\bibfnamefont {F.}~\bibnamefont
  {Hainer}}, \bibinfo {author} {\bibfnamefont {N.}~\bibnamefont {Alagna}},
  \bibinfo {author} {\bibfnamefont {A.}~\bibnamefont {Reddy~Marri}}, \bibinfo
  {author} {\bibfnamefont {T.~J.}\ \bibnamefont {Penfold}}, \bibinfo {author}
  {\bibfnamefont {P.~C.}\ \bibnamefont {Gros}}, \bibinfo {author}
  {\bibfnamefont {S.}~\bibnamefont {Haacke}}, \ and\ \bibinfo {author}
  {\bibfnamefont {T.}~\bibnamefont {Buckup}},\ }\href {\doibase
  10.1021/acs.jpclett.1c01580} {\bibfield  {journal} {\bibinfo  {journal} {The
  Journal of Physical Chemistry Letters}\ }\textbf {\bibinfo {volume} {12}},\
  \bibinfo {pages} {8560} (\bibinfo {year} {2021})}\BibitemShut {NoStop}%
\bibitem [{\citenamefont {Reutzel}\ \emph {et~al.}(2019)\citenamefont
  {Reutzel}, \citenamefont {Li},\ and\ \citenamefont {Petek}}]{Reutzel_PRX19}%
  \BibitemOpen
  \bibfield  {author} {\bibinfo {author} {\bibfnamefont {M.}~\bibnamefont
  {Reutzel}}, \bibinfo {author} {\bibfnamefont {A.}~\bibnamefont {Li}}, \ and\
  \bibinfo {author} {\bibfnamefont {H.}~\bibnamefont {Petek}},\ }\href
  {\doibase 10.1103/PhysRevX.9.011044} {\bibfield  {journal} {\bibinfo
  {journal} {Phys. Rev. X}\ }\textbf {\bibinfo {volume} {9}},\ \bibinfo {pages}
  {011044} (\bibinfo {year} {2019})}\BibitemShut {NoStop}%
\bibitem [{\citenamefont {Bittner}\ and\ \citenamefont
  {Silva}(2014)}]{Bittner_39}%
  \BibitemOpen
  \bibfield  {author} {\bibinfo {author} {\bibfnamefont {E.~R.}\ \bibnamefont
  {Bittner}}\ and\ \bibinfo {author} {\bibfnamefont {C.}~\bibnamefont
  {Silva}},\ }\href {\doibase 10.1038/ncomms4119} {\bibfield  {journal}
  {\bibinfo  {journal} {Nat. Commun.}\ }\textbf {\bibinfo {volume} {5}},\
  \bibinfo {pages} {3119} (\bibinfo {year} {2014})}\BibitemShut {NoStop}%
\bibitem [{\citenamefont {Liedy}\ \emph {et~al.}(2020)\citenamefont {Liedy},
  \citenamefont {Eng}, \citenamefont {McNab}, \citenamefont {Inglis},
  \citenamefont {Penfold}, \citenamefont {Brechin},\ and\ \citenamefont
  {Johansson}}]{Liedy2020}%
  \BibitemOpen
  \bibfield  {author} {\bibinfo {author} {\bibfnamefont {F.}~\bibnamefont
  {Liedy}}, \bibinfo {author} {\bibfnamefont {J.}~\bibnamefont {Eng}}, \bibinfo
  {author} {\bibfnamefont {R.}~\bibnamefont {McNab}}, \bibinfo {author}
  {\bibfnamefont {R.}~\bibnamefont {Inglis}}, \bibinfo {author} {\bibfnamefont
  {T.~J.}\ \bibnamefont {Penfold}}, \bibinfo {author} {\bibfnamefont {E.~K.}\
  \bibnamefont {Brechin}}, \ and\ \bibinfo {author} {\bibfnamefont {J.~O.}\
  \bibnamefont {Johansson}},\ }\href {\doibase 10.1038/s41557-020-0431-6}
  {\bibfield  {journal} {\bibinfo  {journal} {Nature Chemistry}\ }\textbf
  {\bibinfo {volume} {12}},\ \bibinfo {pages} {452} (\bibinfo {year}
  {2020})}\BibitemShut {NoStop}%
\bibitem [{\citenamefont {Rogers}\ \emph {et~al.}(2023)\citenamefont {Rogers},
  \citenamefont {Habib}, \citenamefont {Teobaldi}, \citenamefont {Moorsom},
  \citenamefont {Johansson}, \citenamefont {Hedley}, \citenamefont {Keatley},
  \citenamefont {Hicken}, \citenamefont {Valvidares}, \citenamefont {Gargiani},
  \citenamefont {Alosaimi}, \citenamefont {Poli}, \citenamefont {Ali},
  \citenamefont {Burnell}, \citenamefont {Hickey},\ and\ \citenamefont
  {Cespedes}}]{Rogers_23}%
  \BibitemOpen
  \bibfield  {author} {\bibinfo {author} {\bibfnamefont {M.}~\bibnamefont
  {Rogers}}, \bibinfo {author} {\bibfnamefont {A.}~\bibnamefont {Habib}},
  \bibinfo {author} {\bibfnamefont {G.}~\bibnamefont {Teobaldi}}, \bibinfo
  {author} {\bibfnamefont {T.}~\bibnamefont {Moorsom}}, \bibinfo {author}
  {\bibfnamefont {J.~O.}\ \bibnamefont {Johansson}}, \bibinfo {author}
  {\bibfnamefont {L.}~\bibnamefont {Hedley}}, \bibinfo {author} {\bibfnamefont
  {P.~S.}\ \bibnamefont {Keatley}}, \bibinfo {author} {\bibfnamefont {R.~J.}\
  \bibnamefont {Hicken}}, \bibinfo {author} {\bibfnamefont {M.}~\bibnamefont
  {Valvidares}}, \bibinfo {author} {\bibfnamefont {P.}~\bibnamefont
  {Gargiani}}, \bibinfo {author} {\bibfnamefont {N.}~\bibnamefont {Alosaimi}},
  \bibinfo {author} {\bibfnamefont {E.}~\bibnamefont {Poli}}, \bibinfo {author}
  {\bibfnamefont {M.}~\bibnamefont {Ali}}, \bibinfo {author} {\bibfnamefont
  {G.}~\bibnamefont {Burnell}}, \bibinfo {author} {\bibfnamefont {B.~J.}\
  \bibnamefont {Hickey}}, \ and\ \bibinfo {author} {\bibfnamefont
  {O.}~\bibnamefont {Cespedes}},\ }\href {\doibase
  https://doi.org/10.1002/adfm.202212173} {\bibfield  {journal} {\bibinfo
  {journal} {Advanced Functional Materials}\ }\textbf {\bibinfo {volume}
  {33}},\ \bibinfo {pages} {2212173} (\bibinfo {year} {2023})}\BibitemShut
  {NoStop}%
\bibitem [{\citenamefont {Canton}\ \emph {et~al.}(2023)\citenamefont {Canton},
  \citenamefont {Biednov}, \citenamefont {Pápai}, \citenamefont {Lima},
  \citenamefont {Choi}, \citenamefont {Otte}, \citenamefont {Jiang},
  \citenamefont {Frankenberger}, \citenamefont {Knoll}, \citenamefont {Zalden},
  \citenamefont {Gawelda}, \citenamefont {Rahaman}, \citenamefont {Møller},
  \citenamefont {Milne}, \citenamefont {Gosztola}, \citenamefont {Zheng},
  \citenamefont {Retegan},\ and\ \citenamefont {Khakhulin}}]{Canton_23}%
  \BibitemOpen
  \bibfield  {author} {\bibinfo {author} {\bibfnamefont {S.~E.}\ \bibnamefont
  {Canton}}, \bibinfo {author} {\bibfnamefont {M.}~\bibnamefont {Biednov}},
  \bibinfo {author} {\bibfnamefont {M.}~\bibnamefont {Pápai}}, \bibinfo
  {author} {\bibfnamefont {F.~A.}\ \bibnamefont {Lima}}, \bibinfo {author}
  {\bibfnamefont {T.-K.}\ \bibnamefont {Choi}}, \bibinfo {author}
  {\bibfnamefont {F.}~\bibnamefont {Otte}}, \bibinfo {author} {\bibfnamefont
  {Y.}~\bibnamefont {Jiang}}, \bibinfo {author} {\bibfnamefont
  {P.}~\bibnamefont {Frankenberger}}, \bibinfo {author} {\bibfnamefont
  {M.}~\bibnamefont {Knoll}}, \bibinfo {author} {\bibfnamefont
  {P.}~\bibnamefont {Zalden}}, \bibinfo {author} {\bibfnamefont
  {W.}~\bibnamefont {Gawelda}}, \bibinfo {author} {\bibfnamefont
  {A.}~\bibnamefont {Rahaman}}, \bibinfo {author} {\bibfnamefont {K.~B.}\
  \bibnamefont {Møller}}, \bibinfo {author} {\bibfnamefont {C.}~\bibnamefont
  {Milne}}, \bibinfo {author} {\bibfnamefont {D.~J.}\ \bibnamefont {Gosztola}},
  \bibinfo {author} {\bibfnamefont {K.}~\bibnamefont {Zheng}}, \bibinfo
  {author} {\bibfnamefont {M.}~\bibnamefont {Retegan}}, \ and\ \bibinfo
  {author} {\bibfnamefont {D.}~\bibnamefont {Khakhulin}},\ }\href {\doibase
  https://doi.org/10.1002/advs.202206880} {\bibfield  {journal} {\bibinfo
  {journal} {Advanced Science}\ }\textbf {\bibinfo {volume} {10}},\ \bibinfo
  {pages} {2206880} (\bibinfo {year} {2023})}\BibitemShut {NoStop}%
\bibitem [{\citenamefont {Paulus}\ \emph {et~al.}(2020)\citenamefont {Paulus},
  \citenamefont {Adelman}, \citenamefont {Jamula},\ and\ \citenamefont
  {McCusker}}]{Paulus2020}%
  \BibitemOpen
  \bibfield  {author} {\bibinfo {author} {\bibfnamefont {B.~C.}\ \bibnamefont
  {Paulus}}, \bibinfo {author} {\bibfnamefont {S.~L.}\ \bibnamefont {Adelman}},
  \bibinfo {author} {\bibfnamefont {L.}~\bibnamefont {Jamula}}, \ and\ \bibinfo
  {author} {\bibfnamefont {J.}~\bibnamefont {McCusker}},\ }\href {\doibase
  10.1038/s41586-020-2353-2} {\bibfield  {journal} {\bibinfo  {journal}
  {Nature}\ }\textbf {\bibinfo {volume} {582}},\ \bibinfo {pages} {214}
  (\bibinfo {year} {2020})}\BibitemShut {NoStop}%
\bibitem [{\citenamefont {Coccia}\ and\ \citenamefont
  {Corni}(2019)}]{Coccia_2019}%
  \BibitemOpen
  \bibfield  {author} {\bibinfo {author} {\bibfnamefont {E.}~\bibnamefont
  {Coccia}}\ and\ \bibinfo {author} {\bibfnamefont {S.}~\bibnamefont {Corni}},\
  }\href {\doibase 10.1063/1.5109378} {\bibfield  {journal} {\bibinfo
  {journal} {The Journal of Chemical Physics}\ }\textbf {\bibinfo {volume}
  {151}} (\bibinfo {year} {2019}),\ 10.1063/1.5109378},\ \bibinfo {note}
  {044703}\BibitemShut {NoStop}%
\bibitem [{\citenamefont {Gaynor}\ \emph {et~al.}(2019)\citenamefont {Gaynor},
  \citenamefont {Sandwisch},\ and\ \citenamefont {Khalil}}]{Gaynor2019}%
  \BibitemOpen
  \bibfield  {author} {\bibinfo {author} {\bibfnamefont {J.~D.}\ \bibnamefont
  {Gaynor}}, \bibinfo {author} {\bibfnamefont {J.}~\bibnamefont {Sandwisch}}, \
  and\ \bibinfo {author} {\bibfnamefont {M.}~\bibnamefont {Khalil}},\ }\href
  {\doibase 10.1038/s41467-019-13503-9} {\bibfield  {journal} {\bibinfo
  {journal} {Nature Communications}\ }\textbf {\bibinfo {volume} {10}},\
  \bibinfo {pages} {5621} (\bibinfo {year} {2019})}\BibitemShut {NoStop}%
\bibitem [{\citenamefont {Wang}\ \emph {et~al.}(2016)\citenamefont {Wang},
  \citenamefont {Valkunas}, \citenamefont {Cao}, \citenamefont
  {Whittaker-Brooks},\ and\ \citenamefont {Fleming}}]{Wang_16and1}%
  \BibitemOpen
  \bibfield  {author} {\bibinfo {author} {\bibfnamefont {H.}~\bibnamefont
  {Wang}}, \bibinfo {author} {\bibfnamefont {L.}~\bibnamefont {Valkunas}},
  \bibinfo {author} {\bibfnamefont {T.}~\bibnamefont {Cao}}, \bibinfo {author}
  {\bibfnamefont {L.}~\bibnamefont {Whittaker-Brooks}}, \ and\ \bibinfo
  {author} {\bibfnamefont {G.}~\bibnamefont {Fleming}},\ }\href {\doibase
  10.1021/acs.jpclett.6b01425} {\bibfield  {journal} {\bibinfo  {journal} {J.
  Phys. Chem. Lett.}\ }\textbf {\bibinfo {volume} {7}},\ \bibinfo {pages}
  {3284} (\bibinfo {year} {2016})}\BibitemShut {NoStop}%
\bibitem [{\citenamefont {Liu}\ \emph {et~al.}(2016)\citenamefont {Liu},
  \citenamefont {Xiang}, \citenamefont {Zhang}, \citenamefont {Zhang},
  \citenamefont {Beratan}, \citenamefont {Li},\ and\ \citenamefont
  {Tao}}]{Liu_38}%
  \BibitemOpen
  \bibfield  {author} {\bibinfo {author} {\bibfnamefont {C.}~\bibnamefont
  {Liu}}, \bibinfo {author} {\bibfnamefont {L.}~\bibnamefont {Xiang}}, \bibinfo
  {author} {\bibfnamefont {Y.}~\bibnamefont {Zhang}}, \bibinfo {author}
  {\bibfnamefont {P.}~\bibnamefont {Zhang}}, \bibinfo {author} {\bibfnamefont
  {D.~N.}\ \bibnamefont {Beratan}}, \bibinfo {author} {\bibfnamefont
  {Y.}~\bibnamefont {Li}}, \ and\ \bibinfo {author} {\bibfnamefont
  {N.}~\bibnamefont {Tao}},\ }\href {\doibase 10.1038/nchem.2545} {\bibfield
  {journal} {\bibinfo  {journal} {Nat. Chem.}\ }\textbf {\bibinfo {volume}
  {8}},\ \bibinfo {pages} {941} (\bibinfo {year} {2016})}\BibitemShut {NoStop}%
\bibitem [{\citenamefont {Bian}\ \emph {et~al.}(2020)\citenamefont {Bian},
  \citenamefont {Ma}, \citenamefont {Chen}, \citenamefont {Wei}, \citenamefont
  {Su}, \citenamefont {Buyanova}, \citenamefont {Chen}, \citenamefont
  {Ponseca}, \citenamefont {Linares}, \citenamefont {Karki}, \citenamefont
  {Yartsev},\ and\ \citenamefont {Ingan{\"a}s}}]{Bian2020}%
  \BibitemOpen
  \bibfield  {author} {\bibinfo {author} {\bibfnamefont {Q.}~\bibnamefont
  {Bian}}, \bibinfo {author} {\bibfnamefont {F.}~\bibnamefont {Ma}}, \bibinfo
  {author} {\bibfnamefont {S.}~\bibnamefont {Chen}}, \bibinfo {author}
  {\bibfnamefont {Q.}~\bibnamefont {Wei}}, \bibinfo {author} {\bibfnamefont
  {X.}~\bibnamefont {Su}}, \bibinfo {author} {\bibfnamefont {I.~A.}\
  \bibnamefont {Buyanova}}, \bibinfo {author} {\bibfnamefont {W.~M.}\
  \bibnamefont {Chen}}, \bibinfo {author} {\bibfnamefont {C.~S.}\ \bibnamefont
  {Ponseca}}, \bibinfo {author} {\bibfnamefont {M.}~\bibnamefont {Linares}},
  \bibinfo {author} {\bibfnamefont {K.~J.}\ \bibnamefont {Karki}}, \bibinfo
  {author} {\bibfnamefont {A.}~\bibnamefont {Yartsev}}, \ and\ \bibinfo
  {author} {\bibfnamefont {O.}~\bibnamefont {Ingan{\"a}s}},\ }\href {\doibase
  10.1038/s41467-020-14476-w} {\bibfield  {journal} {\bibinfo  {journal}
  {Nature Communications}\ }\textbf {\bibinfo {volume} {11}},\ \bibinfo {pages}
  {617} (\bibinfo {year} {2020})}\BibitemShut {NoStop}%
\bibitem [{\citenamefont {Dubin}\ \emph {et~al.}(2006)\citenamefont {Dubin},
  \citenamefont {Melet}, \citenamefont {Barisien}, \citenamefont {Grousson},
  \citenamefont {Legrand}, \citenamefont {Schott},\ and\ \citenamefont
  {Voliotis}}]{Dubin_10}%
  \BibitemOpen
  \bibfield  {author} {\bibinfo {author} {\bibfnamefont {F.}~\bibnamefont
  {Dubin}}, \bibinfo {author} {\bibfnamefont {R.}~\bibnamefont {Melet}},
  \bibinfo {author} {\bibfnamefont {T.}~\bibnamefont {Barisien}}, \bibinfo
  {author} {\bibfnamefont {R.}~\bibnamefont {Grousson}}, \bibinfo {author}
  {\bibfnamefont {L.}~\bibnamefont {Legrand}}, \bibinfo {author} {\bibfnamefont
  {M.}~\bibnamefont {Schott}}, \ and\ \bibinfo {author} {\bibfnamefont
  {V.}~\bibnamefont {Voliotis}},\ }\href {\doibase 10.1038/nphys196} {\bibfield
   {journal} {\bibinfo  {journal} {Nat. Phys.}\ }\textbf {\bibinfo {volume}
  {2}},\ \bibinfo {pages} {32} (\bibinfo {year} {2006})}\BibitemShut {NoStop}%
\bibitem [{\citenamefont {Collini}\ and\ \citenamefont
  {Scholes}(2009)}]{Collini_14}%
  \BibitemOpen
  \bibfield  {author} {\bibinfo {author} {\bibfnamefont {E.}~\bibnamefont
  {Collini}}\ and\ \bibinfo {author} {\bibfnamefont {G.~D.}\ \bibnamefont
  {Scholes}},\ }\href {\doibase 10.1126/science.1164016} {\bibfield  {journal}
  {\bibinfo  {journal} {Science}\ }\textbf {\bibinfo {volume} {323}},\ \bibinfo
  {pages} {369} (\bibinfo {year} {2009})}\BibitemShut {NoStop}%
\bibitem [{\citenamefont {Cassette}\ \emph {et~al.}(2015)\citenamefont
  {Cassette}, \citenamefont {Pensack}, \citenamefont {Mahler},\ and\
  \citenamefont {Scholes}}]{Cassette_2}%
  \BibitemOpen
  \bibfield  {author} {\bibinfo {author} {\bibfnamefont {E.}~\bibnamefont
  {Cassette}}, \bibinfo {author} {\bibfnamefont {R.~D.}\ \bibnamefont
  {Pensack}}, \bibinfo {author} {\bibfnamefont {B.}~\bibnamefont {Mahler}}, \
  and\ \bibinfo {author} {\bibfnamefont {G.~D.}\ \bibnamefont {Scholes}},\
  }\href {\doibase 10.1038/ncomms7086} {\bibfield  {journal} {\bibinfo
  {journal} {Nat. Commun.}\ }\textbf {\bibinfo {volume} {6}},\ \bibinfo {pages}
  {6086} (\bibinfo {year} {2015})}\BibitemShut {NoStop}%
\bibitem [{\citenamefont {Scholes}\ and\ \citenamefont
  {Rumbles}(2006)}]{Scholes_25}%
  \BibitemOpen
  \bibfield  {author} {\bibinfo {author} {\bibfnamefont {G.~D.}\ \bibnamefont
  {Scholes}}\ and\ \bibinfo {author} {\bibfnamefont {G.}~\bibnamefont
  {Rumbles}},\ }\href {\doibase 10.1038/nmat1710} {\bibfield  {journal}
  {\bibinfo  {journal} {Nat. Mater.}\ }\textbf {\bibinfo {volume} {5}},\
  \bibinfo {pages} {683} (\bibinfo {year} {2006})}\BibitemShut {NoStop}%
\bibitem [{\citenamefont {Tanimura}(2020)}]{Tanimura_HEOM}%
  \BibitemOpen
  \bibfield  {author} {\bibinfo {author} {\bibfnamefont {Y.}~\bibnamefont
  {Tanimura}},\ }\href {\doibase 10.1063/5.0011599} {\bibfield  {journal}
  {\bibinfo  {journal} {The Journal of Chemical Physics}\ }\textbf {\bibinfo
  {volume} {153}},\ \bibinfo {pages} {020901} (\bibinfo {year}
  {2020})}\BibitemShut {NoStop}%
\bibitem [{\citenamefont {Lambert}\ \emph {et~al.}(2019)\citenamefont
  {Lambert}, \citenamefont {Ahmed}, \citenamefont {Cirio},\ and\ \citenamefont
  {Nori}}]{Lambert2019}%
  \BibitemOpen
  \bibfield  {author} {\bibinfo {author} {\bibfnamefont {N.}~\bibnamefont
  {Lambert}}, \bibinfo {author} {\bibfnamefont {S.}~\bibnamefont {Ahmed}},
  \bibinfo {author} {\bibfnamefont {M.}~\bibnamefont {Cirio}}, \ and\ \bibinfo
  {author} {\bibfnamefont {F.}~\bibnamefont {Nori}},\ }\href {\doibase
  10.1038/s41467-019-11656-1} {\bibfield  {journal} {\bibinfo  {journal}
  {Nature Communications}\ }\textbf {\bibinfo {volume} {10}} (\bibinfo {year}
  {2019}),\ 10.1038/s41467-019-11656-1}\BibitemShut {NoStop}%
\bibitem [{\citenamefont {Egger}\ and\ \citenamefont
  {Mak}(1994)}]{Egger_PRB94}%
  \BibitemOpen
  \bibfield  {author} {\bibinfo {author} {\bibfnamefont {R.}~\bibnamefont
  {Egger}}\ and\ \bibinfo {author} {\bibfnamefont {C.~H.}\ \bibnamefont
  {Mak}},\ }\href {\doibase 10.1103/PhysRevB.50.15210} {\bibfield  {journal}
  {\bibinfo  {journal} {Phys. Rev. B}\ }\textbf {\bibinfo {volume} {50}},\
  \bibinfo {pages} {15210} (\bibinfo {year} {1994})}\BibitemShut {NoStop}%
\bibitem [{\citenamefont {Cao}\ \emph {et~al.}(1996)\citenamefont {Cao},
  \citenamefont {Ungar},\ and\ \citenamefont {Voth}}]{Cao_96}%
  \BibitemOpen
  \bibfield  {author} {\bibinfo {author} {\bibfnamefont {J.}~\bibnamefont
  {Cao}}, \bibinfo {author} {\bibfnamefont {L.~W.}\ \bibnamefont {Ungar}}, \
  and\ \bibinfo {author} {\bibfnamefont {G.~A.}\ \bibnamefont {Voth}},\ }\href
  {\doibase 10.1063/1.471230} {\bibfield  {journal} {\bibinfo  {journal} {The
  Journal of Chemical Physics}\ }\textbf {\bibinfo {volume} {104}},\ \bibinfo
  {pages} {4189} (\bibinfo {year} {1996})}\BibitemShut {NoStop}%
\bibitem [{\citenamefont {Makri}(1995)}]{Makri_95}%
  \BibitemOpen
  \bibfield  {author} {\bibinfo {author} {\bibfnamefont {N.}~\bibnamefont
  {Makri}},\ }\href {\doibase 10.1063/1.531046} {\bibfield  {journal} {\bibinfo
   {journal} {Journal of Mathematical Physics}\ }\textbf {\bibinfo {volume}
  {36}},\ \bibinfo {pages} {2430} (\bibinfo {year} {1995})}\BibitemShut
  {NoStop}%
\bibitem [{\citenamefont {Makri}\ and\ \citenamefont
  {Makarov}(1995)}]{Makri2_95}%
  \BibitemOpen
  \bibfield  {author} {\bibinfo {author} {\bibfnamefont {N.}~\bibnamefont
  {Makri}}\ and\ \bibinfo {author} {\bibfnamefont {D.~E.}\ \bibnamefont
  {Makarov}},\ }\href {\doibase 10.1063/1.469508} {\bibfield  {journal}
  {\bibinfo  {journal} {The Journal of Chemical Physics}\ }\textbf {\bibinfo
  {volume} {102}},\ \bibinfo {pages} {4600} (\bibinfo {year}
  {1995})}\BibitemShut {NoStop}%
\bibitem [{\citenamefont {Zhao}(2023)}]{Zhao_23}%
  \BibitemOpen
  \bibfield  {author} {\bibinfo {author} {\bibfnamefont {Y.}~\bibnamefont
  {Zhao}},\ }\href {\doibase 10.1063/5.0140002} {\bibfield  {journal} {\bibinfo
   {journal} {The Journal of Chemical Physics}\ }\textbf {\bibinfo {volume}
  {158}},\ \bibinfo {pages} {080901} (\bibinfo {year} {2023})}\BibitemShut
  {NoStop}%
\bibitem [{\citenamefont {Liu}\ \emph {et~al.}(2004)\citenamefont {Liu},
  \citenamefont {Yan}, \citenamefont {Wang}, \citenamefont {Kuang},
  \citenamefont {Zhang}, \citenamefont {Gui}, \citenamefont {An},\ and\
  \citenamefont {Chang}}]{Trimer}%
  \BibitemOpen
  \bibfield  {author} {\bibinfo {author} {\bibfnamefont {Z.}~\bibnamefont
  {Liu}}, \bibinfo {author} {\bibfnamefont {H.}~\bibnamefont {Yan}}, \bibinfo
  {author} {\bibfnamefont {K.}~\bibnamefont {Wang}}, \bibinfo {author}
  {\bibfnamefont {T.}~\bibnamefont {Kuang}}, \bibinfo {author} {\bibfnamefont
  {J.}~\bibnamefont {Zhang}}, \bibinfo {author} {\bibfnamefont
  {L.}~\bibnamefont {Gui}}, \bibinfo {author} {\bibfnamefont {X.}~\bibnamefont
  {An}}, \ and\ \bibinfo {author} {\bibfnamefont {W.}~\bibnamefont {Chang}},\
  }\href {\doibase 10.1038/nature02373} {\bibfield  {journal} {\bibinfo
  {journal} {Nature}\ }\textbf {\bibinfo {volume} {428}},\ \bibinfo {pages}
  {287} (\bibinfo {year} {2004})}\BibitemShut {NoStop}%
\bibitem [{\citenamefont {Drop}\ \emph {et~al.}(2014)\citenamefont {Drop},
  \citenamefont {Webber-Birungi}, \citenamefont {Yadav}, \citenamefont
  {Filipowicz-Szymanska}, \citenamefont {Fusetti}, \citenamefont {Boekema},\
  and\ \citenamefont {Croce}}]{Drop}%
  \BibitemOpen
  \bibfield  {author} {\bibinfo {author} {\bibfnamefont {B.}~\bibnamefont
  {Drop}}, \bibinfo {author} {\bibfnamefont {M.}~\bibnamefont
  {Webber-Birungi}}, \bibinfo {author} {\bibfnamefont {S.~K.}\ \bibnamefont
  {Yadav}}, \bibinfo {author} {\bibfnamefont {A.}~\bibnamefont
  {Filipowicz-Szymanska}}, \bibinfo {author} {\bibfnamefont {F.}~\bibnamefont
  {Fusetti}}, \bibinfo {author} {\bibfnamefont {E.~J.}\ \bibnamefont
  {Boekema}}, \ and\ \bibinfo {author} {\bibfnamefont {R.}~\bibnamefont
  {Croce}},\ }\href {\doibase 10.1016/j.bbabio.2013.07.012} {\bibfield
  {journal} {\bibinfo  {journal} {Biochimica et Biophysica Acta ({BBA}) -
  Bioenergetics}\ }\textbf {\bibinfo {volume} {1837}},\ \bibinfo {pages} {63}
  (\bibinfo {year} {2014})}\BibitemShut {NoStop}%
\bibitem [{\citenamefont {Lambrev}\ \emph {et~al.}(2007)\citenamefont
  {Lambrev}, \citenamefont {V{\'{a}}rkonyi}, \citenamefont {Krumova},
  \citenamefont {Kov{\'{a}}cs}, \citenamefont {Miloslavina}, \citenamefont
  {Holzwarth},\ and\ \citenamefont {Garab}}]{Lambrev}%
  \BibitemOpen
  \bibfield  {author} {\bibinfo {author} {\bibfnamefont {P.~H.}\ \bibnamefont
  {Lambrev}}, \bibinfo {author} {\bibfnamefont {Z.}~\bibnamefont
  {V{\'{a}}rkonyi}}, \bibinfo {author} {\bibfnamefont {S.}~\bibnamefont
  {Krumova}}, \bibinfo {author} {\bibfnamefont {L.}~\bibnamefont
  {Kov{\'{a}}cs}}, \bibinfo {author} {\bibfnamefont {Y.}~\bibnamefont
  {Miloslavina}}, \bibinfo {author} {\bibfnamefont {A.~R.}\ \bibnamefont
  {Holzwarth}}, \ and\ \bibinfo {author} {\bibfnamefont {G.}~\bibnamefont
  {Garab}},\ }\href {\doibase 10.1016/j.bbabio.2007.01.010} {\bibfield
  {journal} {\bibinfo  {journal} {Biochimica et Biophysica Acta ({BBA}) -
  Bioenergetics}\ }\textbf {\bibinfo {volume} {1767}},\ \bibinfo {pages} {847}
  (\bibinfo {year} {2007})}\BibitemShut {NoStop}%
\bibitem [{\citenamefont {Duan}\ \emph {et~al.}(2017)\citenamefont {Duan},
  \citenamefont {Prokhorenko}, \citenamefont {Cogdell}, \citenamefont {Ashraf},
  \citenamefont {Stevens}, \citenamefont {Thorwart},\ and\ \citenamefont
  {Miller}}]{PNASnew}%
  \BibitemOpen
  \bibfield  {author} {\bibinfo {author} {\bibfnamefont {H.-G.}\ \bibnamefont
  {Duan}}, \bibinfo {author} {\bibfnamefont {V.~I.}\ \bibnamefont
  {Prokhorenko}}, \bibinfo {author} {\bibfnamefont {R.~J.}\ \bibnamefont
  {Cogdell}}, \bibinfo {author} {\bibfnamefont {K.}~\bibnamefont {Ashraf}},
  \bibinfo {author} {\bibfnamefont {A.~L.}\ \bibnamefont {Stevens}}, \bibinfo
  {author} {\bibfnamefont {M.}~\bibnamefont {Thorwart}}, \ and\ \bibinfo
  {author} {\bibfnamefont {R.~J.~D.}\ \bibnamefont {Miller}},\ }\href {\doibase
  10.1073/pnas.1702261114} {\bibfield  {journal} {\bibinfo  {journal}
  {Proceedings of the National Academy of Sciences}\ }\textbf {\bibinfo
  {volume} {114}},\ \bibinfo {pages} {8493} (\bibinfo {year}
  {2017})}\BibitemShut {NoStop}%
\bibitem [{\citenamefont {Kenrow}\ \emph {et~al.}(1997)\citenamefont {Kenrow},
  \citenamefont {El~Sayed},\ and\ \citenamefont {Stanton}}]{Kenrow}%
  \BibitemOpen
  \bibfield  {author} {\bibinfo {author} {\bibfnamefont {J.~A.}\ \bibnamefont
  {Kenrow}}, \bibinfo {author} {\bibfnamefont {K.}~\bibnamefont {El~Sayed}}, \
  and\ \bibinfo {author} {\bibfnamefont {C.~J.}\ \bibnamefont {Stanton}},\
  }\href {\doibase 10.1103/PhysRevLett.78.4873} {\bibfield  {journal} {\bibinfo
   {journal} {Phys. Rev. Lett.}\ }\textbf {\bibinfo {volume} {78}},\ \bibinfo
  {pages} {4873} (\bibinfo {year} {1997})}\BibitemShut {NoStop}%
\bibitem [{\citenamefont {Li}(2017)}]{Li}%
  \BibitemOpen
  \bibfield  {author} {\bibinfo {author} {\bibfnamefont {C.}~\bibnamefont
  {Li}},\ }\href {\doibase 10.1007/978-981-10-1488-8} {\emph {\bibinfo {title}
  {Nonlinear Optics: Principles and Applications}}}\ (\bibinfo  {publisher}
  {Springer Singapore},\ \bibinfo {address} {Berlin},\ \bibinfo {year} {2017})\
  \bibinfo {note} {p. 25}\BibitemShut {NoStop}%
\bibitem [{\citenamefont {Schr\"oter}(2015)}]{Exc}%
  \BibitemOpen
  \bibfield  {author} {\bibinfo {author} {\bibfnamefont {M.}~\bibnamefont
  {Schr\"oter}},\ }\href {\doibase 10.1007/978-3-658-09282-5} {\emph {\bibinfo
  {title} {Dissipative Exciton Dynamics in Light-Harvesting Complexes}}}\
  (\bibinfo  {publisher} {Springer Spektrum Wiesbaden},\ \bibinfo {address}
  {Berlin},\ \bibinfo {year} {2015})\ \bibinfo {note} {p. 40}\BibitemShut
  {NoStop}%
\bibitem [{\citenamefont {G{\'{o}}mez-Ruiz}\ \emph {et~al.}(2018)\citenamefont
  {G{\'{o}}mez-Ruiz}, \citenamefont {Acevedo}, \citenamefont
  {Rodr{\'{\i}}guez}, \citenamefont {Quiroga},\ and\ \citenamefont
  {Johnson}}]{Johnson}%
  \BibitemOpen
  \bibfield  {author} {\bibinfo {author} {\bibfnamefont {F.~J.}\ \bibnamefont
  {G{\'{o}}mez-Ruiz}}, \bibinfo {author} {\bibfnamefont {O.~L.}\ \bibnamefont
  {Acevedo}}, \bibinfo {author} {\bibfnamefont {F.~J.}\ \bibnamefont
  {Rodr{\'{\i}}guez}}, \bibinfo {author} {\bibfnamefont {L.}~\bibnamefont
  {Quiroga}}, \ and\ \bibinfo {author} {\bibfnamefont {N.~F.}\ \bibnamefont
  {Johnson}},\ }\href {\doibase 10.3389/fphy.2018.00092} {\bibfield  {journal}
  {\bibinfo  {journal} {Frontiers in Physics}\ }\textbf {\bibinfo {volume} {6}}
  (\bibinfo {year} {2018}),\ 10.3389/fphy.2018.00092}\BibitemShut {NoStop}%
\bibitem [{\citenamefont {Rodr\'{\i}guez}\ \emph {et~al.}(2008)\citenamefont
  {Rodr\'{\i}guez}, \citenamefont {Quiroga}, \citenamefont {Tejedor},
  \citenamefont {Mart\'{\i}n}, \citenamefont {Vi\~na},\ and\ \citenamefont
  {Andr\'e}}]{PRBluis}%
  \BibitemOpen
  \bibfield  {author} {\bibinfo {author} {\bibfnamefont {F.~J.}\ \bibnamefont
  {Rodr\'{\i}guez}}, \bibinfo {author} {\bibfnamefont {L.}~\bibnamefont
  {Quiroga}}, \bibinfo {author} {\bibfnamefont {C.}~\bibnamefont {Tejedor}},
  \bibinfo {author} {\bibfnamefont {M.~D.}\ \bibnamefont {Mart\'{\i}n}},
  \bibinfo {author} {\bibfnamefont {L.}~\bibnamefont {Vi\~na}}, \ and\ \bibinfo
  {author} {\bibfnamefont {R.}~\bibnamefont {Andr\'e}},\ }\href {\doibase
  10.1103/PhysRevB.78.035312} {\bibfield  {journal} {\bibinfo  {journal} {Phys.
  Rev. B}\ }\textbf {\bibinfo {volume} {78}},\ \bibinfo {pages} {035312}
  (\bibinfo {year} {2008})}\BibitemShut {NoStop}%
\bibitem [{\citenamefont {Tiwari}\ \emph {et~al.}(2012)\citenamefont {Tiwari},
  \citenamefont {Peters},\ and\ \citenamefont {Jonas}}]{Vivek_22}%
  \BibitemOpen
  \bibfield  {author} {\bibinfo {author} {\bibfnamefont {V.}~\bibnamefont
  {Tiwari}}, \bibinfo {author} {\bibfnamefont {W.~K.}\ \bibnamefont {Peters}},
  \ and\ \bibinfo {author} {\bibfnamefont {D.~M.}\ \bibnamefont {Jonas}},\
  }\href {\doibase 10.1073/pnas.1211157110} {\bibfield  {journal} {\bibinfo
  {journal} {Proceedings of the National Academy of Sciences}\ }\textbf
  {\bibinfo {volume} {110}},\ \bibinfo {pages} {1203} (\bibinfo {year}
  {2012})}\BibitemShut {NoStop}%
\bibitem [{\citenamefont {Acevedo}\ \emph {et~al.}(2014)\citenamefont
  {Acevedo}, \citenamefont {Quiroga}, \citenamefont {Rodr{\'{\i}}guez},\ and\
  \citenamefont {Johnson}}]{Acevedo2014PRL}%
  \BibitemOpen
  \bibfield  {author} {\bibinfo {author} {\bibfnamefont {O.}~\bibnamefont
  {Acevedo}}, \bibinfo {author} {\bibfnamefont {L.}~\bibnamefont {Quiroga}},
  \bibinfo {author} {\bibfnamefont {F.}~\bibnamefont {Rodr{\'{\i}}guez}}, \
  and\ \bibinfo {author} {\bibfnamefont {N.}~\bibnamefont {Johnson}},\ }\href
  {\doibase 10.1103/physrevlett.112.030403} {\bibfield  {journal} {\bibinfo
  {journal} {Physical Review Letters}\ }\textbf {\bibinfo {volume} {112}}
  (\bibinfo {year} {2014}),\ 10.1103/physrevlett.112.030403}\BibitemShut
  {NoStop}%
\bibitem [{\citenamefont {Lee}\ and\ \citenamefont
  {Johnson}(2007)}]{Lee2007_a}%
  \BibitemOpen
  \bibfield  {author} {\bibinfo {author} {\bibfnamefont {C.~F.}\ \bibnamefont
  {Lee}}\ and\ \bibinfo {author} {\bibfnamefont {N.~F.}\ \bibnamefont
  {Johnson}},\ }\href {\doibase 10.1209/0295-5075/81/37004} {\bibfield
  {journal} {\bibinfo  {journal} {{EPL} (Europhysics Letters)}\ }\textbf
  {\bibinfo {volume} {81}},\ \bibinfo {pages} {37004} (\bibinfo {year}
  {2007})}\BibitemShut {NoStop}%
\bibitem [{\citenamefont {Jarrett}\ \emph {et~al.}(2006)\citenamefont
  {Jarrett}, \citenamefont {Lee},\ and\ \citenamefont {Johnson}}]{Jarrett_b}%
  \BibitemOpen
  \bibfield  {author} {\bibinfo {author} {\bibfnamefont {T.~C.}\ \bibnamefont
  {Jarrett}}, \bibinfo {author} {\bibfnamefont {C.~F.}\ \bibnamefont {Lee}}, \
  and\ \bibinfo {author} {\bibfnamefont {N.~F.}\ \bibnamefont {Johnson}},\
  }\href {\doibase 10.1103/PhysRevB.74.121301} {\bibfield  {journal} {\bibinfo
  {journal} {Phys. Rev. B}\ }\textbf {\bibinfo {volume} {74}},\ \bibinfo
  {pages} {121301} (\bibinfo {year} {2006})}\BibitemShut {NoStop}%
\bibitem [{\citenamefont {Lee}\ and\ \citenamefont {Johnson}(2004)}]{Lee_c}%
  \BibitemOpen
  \bibfield  {author} {\bibinfo {author} {\bibfnamefont {C.~F.}\ \bibnamefont
  {Lee}}\ and\ \bibinfo {author} {\bibfnamefont {N.~F.}\ \bibnamefont
  {Johnson}},\ }\href {\doibase 10.1103/PhysRevLett.93.083001} {\bibfield
  {journal} {\bibinfo  {journal} {Phys. Rev. Lett.}\ }\textbf {\bibinfo
  {volume} {93}},\ \bibinfo {pages} {083001} (\bibinfo {year}
  {2004})}\BibitemShut {NoStop}%
\bibitem [{\citenamefont {Jarrett}\ \emph {et~al.}(2007)\citenamefont
  {Jarrett}, \citenamefont {Olaya-Castro},\ and\ \citenamefont
  {Johnson}}]{Jarrett_d}%
  \BibitemOpen
  \bibfield  {author} {\bibinfo {author} {\bibfnamefont {T.~C.}\ \bibnamefont
  {Jarrett}}, \bibinfo {author} {\bibfnamefont {A.}~\bibnamefont
  {Olaya-Castro}}, \ and\ \bibinfo {author} {\bibfnamefont {N.~F.}\
  \bibnamefont {Johnson}},\ }\href {\doibase 10.1209/0295-5075/77/34001}
  {\bibfield  {journal} {\bibinfo  {journal} {Europhysics Letters ({EPL})}\
  }\textbf {\bibinfo {volume} {77}},\ \bibinfo {pages} {34001} (\bibinfo {year}
  {2007})}\BibitemShut {NoStop}%
\bibitem [{\citenamefont {Acevedo}\ \emph
  {et~al.}(2015{\natexlab{a}})\citenamefont {Acevedo}, \citenamefont {Quiroga},
  \citenamefont {Rodr{\'{\i}}guez},\ and\ \citenamefont
  {Johnson}}]{Acevedo2015NJP}%
  \BibitemOpen
  \bibfield  {author} {\bibinfo {author} {\bibfnamefont {O.~L.}\ \bibnamefont
  {Acevedo}}, \bibinfo {author} {\bibfnamefont {L.}~\bibnamefont {Quiroga}},
  \bibinfo {author} {\bibfnamefont {F.~J.}\ \bibnamefont {Rodr{\'{\i}}guez}}, \
  and\ \bibinfo {author} {\bibfnamefont {N.~F.}\ \bibnamefont {Johnson}},\
  }\href {\doibase 10.1088/1367-2630/17/9/093005} {\bibfield  {journal}
  {\bibinfo  {journal} {New Journal of Physics}\ }\textbf {\bibinfo {volume}
  {17}},\ \bibinfo {pages} {093005} (\bibinfo {year}
  {2015}{\natexlab{a}})}\BibitemShut {NoStop}%
\bibitem [{\citenamefont {Acevedo}\ \emph
  {et~al.}(2015{\natexlab{b}})\citenamefont {Acevedo}, \citenamefont {Quiroga},
  \citenamefont {Rodr\'{\i}guez},\ and\ \citenamefont
  {Johnson}}]{AcevedoPRA2015}%
  \BibitemOpen
  \bibfield  {author} {\bibinfo {author} {\bibfnamefont {O.~L.}\ \bibnamefont
  {Acevedo}}, \bibinfo {author} {\bibfnamefont {L.}~\bibnamefont {Quiroga}},
  \bibinfo {author} {\bibfnamefont {F.~J.}\ \bibnamefont {Rodr\'{\i}guez}}, \
  and\ \bibinfo {author} {\bibfnamefont {N.~F.}\ \bibnamefont {Johnson}},\
  }\href {\doibase 10.1103/PhysRevA.92.032330} {\bibfield  {journal} {\bibinfo
  {journal} {Phys. Rev. A}\ }\textbf {\bibinfo {volume} {92}},\ \bibinfo
  {pages} {032330} (\bibinfo {year} {2015}{\natexlab{b}})}\BibitemShut
  {NoStop}%
\bibitem [{\citenamefont {G{\'{o}}mez-Ruiz}\ \emph {et~al.}(2016)\citenamefont
  {G{\'{o}}mez-Ruiz}, \citenamefont {Acevedo}, \citenamefont {Quiroga},
  \citenamefont {Rodr{\'{\i}}guez},\ and\ \citenamefont
  {Johnson}}]{GomezEnt2016}%
  \BibitemOpen
  \bibfield  {author} {\bibinfo {author} {\bibfnamefont {F.}~\bibnamefont
  {G{\'{o}}mez-Ruiz}}, \bibinfo {author} {\bibfnamefont {O.}~\bibnamefont
  {Acevedo}}, \bibinfo {author} {\bibfnamefont {L.}~\bibnamefont {Quiroga}},
  \bibinfo {author} {\bibfnamefont {F.}~\bibnamefont {Rodr{\'{\i}}guez}}, \
  and\ \bibinfo {author} {\bibfnamefont {N.}~\bibnamefont {Johnson}},\ }\href
  {\doibase 10.3390/e18090319} {\bibfield  {journal} {\bibinfo  {journal}
  {Entropy}\ }\textbf {\bibinfo {volume} {18}},\ \bibinfo {pages} {319}
  (\bibinfo {year} {2016})}\BibitemShut {NoStop}%
\bibitem [{\citenamefont {G{\'{o}}mez-Ruiz}\ \emph {et~al.}(2017)\citenamefont
  {G{\'{o}}mez-Ruiz}, \citenamefont {Mendoza-Arenas}, \citenamefont {Acevedo},
  \citenamefont {Rodr{\'{\i}}guez}, \citenamefont {Quiroga},\ and\
  \citenamefont {Johnson}}]{GomezJPB2018}%
  \BibitemOpen
  \bibfield  {author} {\bibinfo {author} {\bibfnamefont {F.~J.}\ \bibnamefont
  {G{\'{o}}mez-Ruiz}}, \bibinfo {author} {\bibfnamefont {J.~J.}\ \bibnamefont
  {Mendoza-Arenas}}, \bibinfo {author} {\bibfnamefont {O.~L.}\ \bibnamefont
  {Acevedo}}, \bibinfo {author} {\bibfnamefont {F.~J.}\ \bibnamefont
  {Rodr{\'{\i}}guez}}, \bibinfo {author} {\bibfnamefont {L.}~\bibnamefont
  {Quiroga}}, \ and\ \bibinfo {author} {\bibfnamefont {N.~F.}\ \bibnamefont
  {Johnson}},\ }\href {\doibase 10.1088/1361-6455/aa9a92} {\bibfield  {journal}
  {\bibinfo  {journal} {Journal of Physics B: Atomic, Molecular and Optical
  Physics}\ }\textbf {\bibinfo {volume} {51}},\ \bibinfo {pages} {024001}
  (\bibinfo {year} {2017})}\BibitemShut {NoStop}%
\bibitem [{\citenamefont {M\'endez-C\'ordoba}\ \emph
  {et~al.}(2020)\citenamefont {M\'endez-C\'ordoba}, \citenamefont
  {Mendoza-Arenas}, \citenamefont {G\'omez-Ruiz}, \citenamefont
  {Rodr\'{\i}guez}, \citenamefont {Tejedor},\ and\ \citenamefont
  {Quiroga}}]{Fabio_2020}%
  \BibitemOpen
  \bibfield  {author} {\bibinfo {author} {\bibfnamefont {F.~P.~M.}\
  \bibnamefont {M\'endez-C\'ordoba}}, \bibinfo {author} {\bibfnamefont {J.~J.}\
  \bibnamefont {Mendoza-Arenas}}, \bibinfo {author} {\bibfnamefont {F.~J.}\
  \bibnamefont {G\'omez-Ruiz}}, \bibinfo {author} {\bibfnamefont {F.~J.}\
  \bibnamefont {Rodr\'{\i}guez}}, \bibinfo {author} {\bibfnamefont
  {C.}~\bibnamefont {Tejedor}}, \ and\ \bibinfo {author} {\bibfnamefont
  {L.}~\bibnamefont {Quiroga}},\ }\href {\doibase
  10.1103/PhysRevResearch.2.043264} {\bibfield  {journal} {\bibinfo  {journal}
  {Phys. Rev. Res.}\ }\textbf {\bibinfo {volume} {2}},\ \bibinfo {pages}
  {043264} (\bibinfo {year} {2020})}\BibitemShut {NoStop}%
\bibitem [{\citenamefont {M\'endez-C\'ordoba}\ \emph
  {et~al.}(2023)\citenamefont {M\'endez-C\'ordoba}, \citenamefont
  {Rodr\'{\i}guez}, \citenamefont {Tejedor},\ and\ \citenamefont
  {Quiroga}}]{Fabio_2023}%
  \BibitemOpen
  \bibfield  {author} {\bibinfo {author} {\bibfnamefont {F.~P.~M.}\
  \bibnamefont {M\'endez-C\'ordoba}}, \bibinfo {author} {\bibfnamefont {F.~J.}\
  \bibnamefont {Rodr\'{\i}guez}}, \bibinfo {author} {\bibfnamefont
  {C.}~\bibnamefont {Tejedor}}, \ and\ \bibinfo {author} {\bibfnamefont
  {L.}~\bibnamefont {Quiroga}},\ }\href {\doibase 10.1103/PhysRevB.107.125104}
  {\bibfield  {journal} {\bibinfo  {journal} {Phys. Rev. B}\ }\textbf {\bibinfo
  {volume} {107}},\ \bibinfo {pages} {125104} (\bibinfo {year}
  {2023})}\BibitemShut {NoStop}%
\bibitem [{\citenamefont {Tarantelli}\ and\ \citenamefont
  {Vicari}(2022)}]{Tarantelli_PRB22}%
  \BibitemOpen
  \bibfield  {author} {\bibinfo {author} {\bibfnamefont {F.}~\bibnamefont
  {Tarantelli}}\ and\ \bibinfo {author} {\bibfnamefont {E.}~\bibnamefont
  {Vicari}},\ }\href {\doibase 10.1103/PhysRevB.105.235124} {\bibfield
  {journal} {\bibinfo  {journal} {Phys. Rev. B}\ }\textbf {\bibinfo {volume}
  {105}},\ \bibinfo {pages} {235124} (\bibinfo {year} {2022})}\BibitemShut
  {NoStop}%
\bibitem [{\citenamefont {De~Franco}\ and\ \citenamefont
  {Vicari}(2023)}]{Franco_PRB23}%
  \BibitemOpen
  \bibfield  {author} {\bibinfo {author} {\bibfnamefont {F.}~\bibnamefont
  {De~Franco}}\ and\ \bibinfo {author} {\bibfnamefont {E.}~\bibnamefont
  {Vicari}},\ }\href {\doibase 10.1103/PhysRevB.107.115175} {\bibfield
  {journal} {\bibinfo  {journal} {Phys. Rev. B}\ }\textbf {\bibinfo {volume}
  {107}},\ \bibinfo {pages} {115175} (\bibinfo {year} {2023})}\BibitemShut
  {NoStop}%
\bibitem [{\citenamefont {Rey}\ \emph {et~al.}(2007)\citenamefont {Rey},
  \citenamefont {Jiang},\ and\ \citenamefont {Lukin}}]{Rey2007}%
  \BibitemOpen
  \bibfield  {author} {\bibinfo {author} {\bibfnamefont {A.~M.}\ \bibnamefont
  {Rey}}, \bibinfo {author} {\bibfnamefont {L.}~\bibnamefont {Jiang}}, \ and\
  \bibinfo {author} {\bibfnamefont {M.~D.}\ \bibnamefont {Lukin}},\ }\href
  {\doibase 10.1103/PhysRevA.76.053617} {\bibfield  {journal} {\bibinfo
  {journal} {Phys. Rev. A}\ }\textbf {\bibinfo {volume} {76}},\ \bibinfo
  {pages} {053617} (\bibinfo {year} {2007})}\BibitemShut {NoStop}%
\bibitem [{\citenamefont {Hardal}\ and\ \citenamefont {\"{O}zg\"{u}r
  E.~M\"{u}stecapl{\i}o{\u{g}}lu}(2015)}]{Hardal_CRP2015}%
  \BibitemOpen
  \bibfield  {author} {\bibinfo {author} {\bibfnamefont {A.~U.~C.}\
  \bibnamefont {Hardal}}\ and\ \bibinfo {author} {\bibnamefont {\"{O}zg\"{u}r
  E.~M\"{u}stecapl{\i}o{\u{g}}lu}},\ }\href {\doibase 10.1038/srep12953}
  {\bibfield  {journal} {\bibinfo  {journal} {Scientific Reports}\ }\textbf
  {\bibinfo {volume} {5}} (\bibinfo {year} {2015}),\
  10.1038/srep12953}\BibitemShut {NoStop}%
\bibitem [{\citenamefont {Niedenzu}\ \emph {et~al.}(2015)\citenamefont
  {Niedenzu}, \citenamefont {Gelbwaser-Klimovsky},\ and\ \citenamefont
  {Kurizki}}]{NiedenzuPRE2015}%
  \BibitemOpen
  \bibfield  {author} {\bibinfo {author} {\bibfnamefont {W.}~\bibnamefont
  {Niedenzu}}, \bibinfo {author} {\bibfnamefont {D.}~\bibnamefont
  {Gelbwaser-Klimovsky}}, \ and\ \bibinfo {author} {\bibfnamefont
  {G.}~\bibnamefont {Kurizki}},\ }\href {\doibase 10.1103/PhysRevE.92.042123}
  {\bibfield  {journal} {\bibinfo  {journal} {Phys. Rev. E}\ }\textbf {\bibinfo
  {volume} {92}},\ \bibinfo {pages} {042123} (\bibinfo {year}
  {2015})}\BibitemShut {NoStop}%
\bibitem [{\citenamefont {Viehmann}\ \emph {et~al.}(2011)\citenamefont
  {Viehmann}, \citenamefont {von Delft},\ and\ \citenamefont
  {Marquardt}}]{Marquadt2011}%
  \BibitemOpen
  \bibfield  {author} {\bibinfo {author} {\bibfnamefont {O.}~\bibnamefont
  {Viehmann}}, \bibinfo {author} {\bibfnamefont {J.}~\bibnamefont {von Delft}},
  \ and\ \bibinfo {author} {\bibfnamefont {F.}~\bibnamefont {Marquardt}},\
  }\href {\doibase 10.1103/PhysRevLett.107.113602} {\bibfield  {journal}
  {\bibinfo  {journal} {Phys. Rev. Lett.}\ }\textbf {\bibinfo {volume} {107}},\
  \bibinfo {pages} {113602} (\bibinfo {year} {2011})}\BibitemShut {NoStop}%
\bibitem [{\citenamefont {Reslen}\ \emph {et~al.}(2005)\citenamefont {Reslen},
  \citenamefont {Quiroga},\ and\ \citenamefont {Johnson}}]{Reslen2005epl}%
  \BibitemOpen
  \bibfield  {author} {\bibinfo {author} {\bibfnamefont {J.}~\bibnamefont
  {Reslen}}, \bibinfo {author} {\bibfnamefont {L.}~\bibnamefont {Quiroga}}, \
  and\ \bibinfo {author} {\bibfnamefont {N.~F.}\ \bibnamefont {Johnson}},\
  }\href {\doibase 10.1209/epl/i2004-10313-4} {\bibfield  {journal} {\bibinfo
  {journal} {Europhysics Letters ({EPL})}\ }\textbf {\bibinfo {volume} {69}},\
  \bibinfo {pages} {8} (\bibinfo {year} {2005})}\BibitemShut {NoStop}%
\bibitem [{\citenamefont {Scully}\ and\ \citenamefont
  {Zubairy}(1997)}]{Scully1997}%
  \BibitemOpen
  \bibfield  {author} {\bibinfo {author} {\bibfnamefont {M.~O.}\ \bibnamefont
  {Scully}}\ and\ \bibinfo {author} {\bibfnamefont {M.~S.}\ \bibnamefont
  {Zubairy}},\ }\href {\doibase 10.1017/cbo9780511813993} {\emph {\bibinfo
  {title} {Quantum Optics}}}\ (\bibinfo  {publisher} {Cambridge University
  Press},\ \bibinfo {year} {1997})\BibitemShut {NoStop}%
\bibitem [{\citenamefont {Vidal}\ and\ \citenamefont
  {Werner}(2002)}]{Vidalneg}%
  \BibitemOpen
  \bibfield  {author} {\bibinfo {author} {\bibfnamefont {G.}~\bibnamefont
  {Vidal}}\ and\ \bibinfo {author} {\bibfnamefont {R.~F.}\ \bibnamefont
  {Werner}},\ }\href {\doibase 10.1103/PhysRevA.65.032314} {\bibfield
  {journal} {\bibinfo  {journal} {Phys. Rev. A}\ }\textbf {\bibinfo {volume}
  {65}},\ \bibinfo {pages} {032314} (\bibinfo {year} {2002})}\BibitemShut
  {NoStop}%
\bibitem [{\citenamefont {Breuer}\ and\ \citenamefont
  {Petruccione}(2002)}]{breuer2002theory}%
  \BibitemOpen
  \bibfield  {author} {\bibinfo {author} {\bibfnamefont {H.}~\bibnamefont
  {Breuer}}\ and\ \bibinfo {author} {\bibfnamefont {F.}~\bibnamefont
  {Petruccione}},\ }\href {https://books.google.es/books?id=0Yx5VzaMYm8C}
  {\emph {\bibinfo {title} {The Theory of Open Quantum Systems}}}\ (\bibinfo
  {publisher} {Oxford University Press},\ \bibinfo {year} {2002})\BibitemShut
  {NoStop}%
\bibitem [{\citenamefont {Shim}\ \emph {et~al.}(2012)\citenamefont {Shim},
  \citenamefont {Rebentrost}, \citenamefont {Valleau},\ and\ \citenamefont
  {Aspuru-Guzik}}]{Sangwoo}%
  \BibitemOpen
  \bibfield  {author} {\bibinfo {author} {\bibfnamefont {S.}~\bibnamefont
  {Shim}}, \bibinfo {author} {\bibfnamefont {P.}~\bibnamefont {Rebentrost}},
  \bibinfo {author} {\bibfnamefont {S.}~\bibnamefont {Valleau}}, \ and\
  \bibinfo {author} {\bibfnamefont {A.}~\bibnamefont {Aspuru-Guzik}},\ }\href
  {\doibase https://doi.org/10.1016/j.bpj.2011.12.021} {\bibfield  {journal}
  {\bibinfo  {journal} {Biophysical Journal}\ }\textbf {\bibinfo {volume}
  {102}},\ \bibinfo {pages} {649} (\bibinfo {year} {2012})}\BibitemShut
  {NoStop}%
\end{thebibliography}%
\bibliographystyle{apsrev4-1}
\end{document}